\input harvmac

\input amssym

\def\unit{\relax{\rm 1\kern-.26em I}}
\def\nada{\relax{\rm 0\kern-.30em l}}
\def\tilde{\widetilde}

\def\alphadot{{\dot \alpha}}



\noblackbox
\def\IL{\relax{\rm I\kern-.18em L}}
\def\IH{\relax{\rm I\kern-.18em H}}
\def\IR{\relax{\rm I\kern-.18em R}}
\def\IC{\relax\hbox{$\inbar\kern-.3em{\rm C}$}}
\def\IZ{\relax\ifmmode\mathchoice
{\hbox{\cmss Z\kern-.4em Z}}{\hbox{\cmss Z\kern-.4em Z}} {\lower.9pt\hbox{\cmsss Z\kern-.4em Z}}
{\lower1.2pt\hbox{\cmsss Z\kern-.4em Z}}\else{\cmss Z\kern-.4em Z}\fi}

\def\CR {{\cal R}}

\def\CJ {{\cal J}}

\def\CL {{\cal L}}

\def\CO {{\cal O}}

\def\CH {{\cal H}}

\def\CA{{\cal A}}


\def\CO {{\cal O}}

\def\Xnl{X_{NL}}
\def\CH{{\cal H}}

\def\alphadot{{\dot \alpha}}

\def\ibar{{\bar i}}

\font\manual=manfnt \def\dbend{\lower3.5pt\hbox{\manual\char127}}

\def\IZ{\relax\ifmmode\mathchoice
{\hbox{\cmss Z\kern-.4em Z}}{\hbox{\cmss Z\kern-.4em Z}} {\lower.9pt\hbox{\cmsss Z\kern-.4em Z}}
{\lower1.2pt\hbox{\cmsss Z\kern-.4em Z}}\else{\cmss Z\kern-.4em Z}\fi}

\def\bar{\overline}

\def\CH{{\cal H}}

\def\psif{\psi_{\phi}}
\def\pa{\partial}

\def\rt2{\sqrt{2}}
\def\irt2{{1\over\sqrt{2}}}

\def\hat{\widehat}
\def\slashchar#1{\setbox0=\hbox{$#1$}           
   \dimen0=\wd0                                 
   \setbox1=\hbox{/} \dimen1=\wd1               
   \ifdim\dimen0>\dimen1                        
      \rlap{\hbox to \dimen0{\hfil/\hfil}}      
      #1                                        
   \else                                        
      \rlap{\hbox to \dimen1{\hfil$#1$\hfil}}   
      /                                         
   \fi}

\def\foursqr#1#2{{\vcenter{\vbox{
    \hrule height.#2pt
    \hbox{\vrule width.#2pt height#1pt \kern#1pt
    \vrule width.#2pt}
    \hrule height.#2pt
    \hrule height.#2pt
    \hbox{\vrule width.#2pt height#1pt \kern#1pt
    \vrule width.#2pt}
    \hrule height.#2pt
        \hrule height.#2pt
    \hbox{\vrule width.#2pt height#1pt \kern#1pt
    \vrule width.#2pt}
    \hrule height.#2pt
        \hrule height.#2pt
    \hbox{\vrule width.#2pt height#1pt \kern#1pt
    \vrule width.#2pt}
    \hrule height.#2pt}}}}
\def\psqr#1#2{{\vcenter{\vbox{\hrule height.#2pt
    \hbox{\vrule width.#2pt height#1pt \kern#1pt
    \vrule width.#2pt}
    \hrule height.#2pt \hrule height.#2pt
    \hbox{\vrule width.#2pt height#1pt \kern#1pt
    \vrule width.#2pt}
    \hrule height.#2pt}}}}
\def\sqr#1#2{{\vcenter{\vbox{\hrule height.#2pt
    \hbox{\vrule width.#2pt height#1pt \kern#1pt
    \vrule width.#2pt}
    \hrule height.#2pt}}}}

\def\figin{\epsfcheck\figin}\def\figins{\epsfcheck\figins}
\def\epsfcheck{\ifx\epsfbox\UnDeFiNeD
\message{(NO epsf.tex, FIGURES WILL BE IGNORED)}
\gdef\figin##1{\vskip2in}\gdef\figins##1{\hskip.5in}
\else\message{(FIGURES WILL BE INCLUDED)}%
\gdef\figin##1{##1}\gdef\figins##1{##1}\fi}
\def\DefWarn#1{}
\def\figinsert{\goodbreak\midinsert}
\def\ifig#1#2#3{\DefWarn#1\xdef#1{fig.~\the\figno}
\writedef{#1\leftbracket fig.\noexpand~\the\figno}%
\figinsert\figin{\centerline{#3}}\medskip\centerline{\vbox{\baselineskip12pt \advance\hsize by
-1truein\noindent\footnotefont{\bf Fig.~\the\figno:\ } \it#2}}
\bigskip\endinsert\global\advance\figno by1}


\lref\SamuelUH{
  S.~Samuel and J.~Wess,
  ``A Superfield Formulation Of The Nonlinear Realization Of Supersymmetry And
  Its Coupling To Supergravity,''
  Nucl.\ Phys.\  B {\bf 221}, 153 (1983).
}

\lref\WessCP{
  J.~Wess and J.~Bagger,
  ``Supersymmetry and supergravity,''
{\it  Princeton, USA: Univ. Pr. (1992) 259 p}
}

\lref\BaggerGM{
  J.~A.~Bagger and A.~F.~Falk,
  ``Decoupling and Destabilizing in Spontaneously Broken Supersymmetry,''
  Phys.\ Rev.\  D {\bf 76}, 105026 (2007)
  [arXiv:0708.3364 [hep-ph]].
}

\lref\RocekNB{
  M.~Rocek,
  ``Linearizing The Volkov-Akulov Model,''
  Phys.\ Rev.\ Lett.\  {\bf 41}, 451 (1978).
}

\lref\peskin{M.~Peskin, unpublished.}

\lref\GirardelloWZ{
  L.~Girardello and M.~T.~Grisaru,
  ``Soft Breaking Of Supersymmetry,''
  Nucl.\ Phys.\  B {\bf 194}, 65 (1982).
}

\lref\IvanovMY{
  E.~A.~Ivanov and A.~A.~Kapustnikov,
  ``Relation Between Linear And Nonlinear Realizations Of Supersymmetry,''
  JINR-E2-10765, Jun 1977
}

\lref\IvanovMX{
  E.~A.~Ivanov and A.~A.~Kapustnikov,
  ``General Relationship Between Linear And Nonlinear Realizations Of
  Supersymmetry,''
  J.\ Phys.\ A  {\bf 11}, 2375 (1978).
}
\lref\IvanovBPA{
  E.~A.~Ivanov and A.~A.~Kapustnikov,
  ``The Nonlinear Realization Structure Of Models With Spontaneously Broken
  Supersymmetry,''
  J.\ Phys.\ G {\bf 8}, 167 (1982).
}

\lref\AffleckVC{
  I.~Affleck, M.~Dine and N.~Seiberg,
  ``Dynamical Supersymmetry Breaking In Chiral Theories,''
  Phys.\ Lett.\  B {\bf 137}, 187 (1984).
}

\lref\SamuelUH{
  S.~Samuel and J.~Wess,
  ``A Superfield Formulation Of The Nonlinear Realization Of Supersymmetry And
  Its Coupling To Supergravity,''
  Nucl.\ Phys.\  B {\bf 221}, 153 (1983).
}

\lref\WessCP{
  J.~Wess and J.~Bagger,
  ``Supersymmetry and supergravity,''
{\it  Princeton, USA: Univ. Pr. (1992) 259 p}
}

\lref\RocekNB{
  M.~Rocek,
  ``Linearizing The Volkov-Akulov Model,''
  Phys.\ Rev.\ Lett.\  {\bf 41}, 451 (1978).
}

\lref\FerraraPZ{
  S.~Ferrara and B.~Zumino,
  ``Transformation Properties Of The Supercurrent,''
  Nucl.\ Phys.\  B {\bf 87}, 207 (1975).
}

\lref\BrignolePE{
  A.~Brignole, F.~Feruglio and F.~Zwirner,
  ``On the effective interactions of a light gravitino with matter  fermions,''
  JHEP {\bf 9711}, 001 (1997)
  [arXiv:hep-th/9709111].
}

\lref\LutyNP{
  M.~A.~Luty and E.~Ponton,
  ``Effective Lagrangians and light gravitino phenomenology,''
  Phys.\ Rev.\  D {\bf 57}, 4167 (1998)
  [arXiv:hep-ph/9706268].
}

\lref\BrignoleSK{
  A.~Brignole, F.~Feruglio and F.~Zwirner,
  ``Signals of a superlight gravitino at e+ e- colliders when the other
  superparticles are heavy,''
  Nucl.\ Phys.\  B {\bf 516}, 13 (1998)
  [Erratum-ibid.\  B {\bf 555}, 653 (1999)]
  [arXiv:hep-ph/9711516].
}

\lref\DineXI{
  M.~Dine, N.~Seiberg and S.~Thomas,
  ``Higgs Physics as a Window Beyond the MSSM (BMSSM),''
  Phys.\ Rev.\  D {\bf 76}, 095004 (2007)
  [arXiv:0707.0005 [hep-ph]].
}

\lref\KomargodskiPC{
  Z.~Komargodski and N.~Seiberg,
  ``Comments on the Fayet-Iliopoulos Term in Field Theory and Supergravity,''
  JHEP {\bf 0906}, 007 (2009)
  [arXiv:0904.1159 [hep-th]].
}

\lref\BrignoleME{
  A.~Brignole, F.~Feruglio, M.~L.~Mangano and F.~Zwirner,
  Nucl.\ Phys.\  B {\bf 526}, 136 (1998)
  [Erratum-ibid.\  B {\bf 582}, 759 (2000)]
  [arXiv:hep-ph/9801329].
}

\lref\ClarkXJ{
  T.~E.~Clark and S.~T.~Love,
  ``Nonlinear realization of supersymmetry and superconformal symmetry,''
  Phys.\ Rev.\  D {\bf 70}, 105011 (2004)
  [arXiv:hep-th/0404162].
}

\lref\ClarkAA{
  T.~E.~Clark, T.~Lee, S.~T.~Love and G.~H.~Wu,
  ``On the interactions of light gravitinos,''
  Phys.\ Rev.\  D {\bf 57}, 5912 (1998)
  [arXiv:hep-ph/9712353].
}

\lref\MeadeWD{
  P.~Meade, N.~Seiberg and D.~Shih,
  ``General Gauge Mediation,''
  arXiv:0801.3278 [hep-ph].
}

\lref\AffleckVC{
  I.~Affleck, M.~Dine and N.~Seiberg,
  ``Dynamical Supersymmetry Breaking In Chiral Theories,''
  Phys.\ Lett.\  B {\bf 137}, 187 (1984).
}

\lref\ClarkBG{
  T.~E.~Clark and S.~T.~Love,
  ``The Supercurrent in supersymmetric field theories,''
  Int.\ J.\ Mod.\ Phys.\  A {\bf 11}, 2807 (1996)
  [arXiv:hep-th/9506145].
}

\lref\VolkovIX{
  D.~V.~Volkov and V.~P.~Akulov,
  ``Is the Neutrino a Goldstone Particle?,''
  Phys.\ Lett.\  B {\bf 46}, 109 (1973).
}

\lref\BrignoleFN{
  A.~Brignole, F.~Feruglio and F.~Zwirner,
  ``Aspects of spontaneously broken N = 1 global supersymmetry in the  presence
  of gauge interactions,''
  Nucl.\ Phys.\  B {\bf 501}, 332 (1997)
  [arXiv:hep-ph/9703286].
}

\lref\LindstromKQ{
  U.~Lindstrom and M.~Rocek,
  ``Constrained Local Superfields,''
  Phys.\ Rev.\  D {\bf 19}, 2300 (1979).
}

\lref\AffleckMF{
  I.~Affleck, M.~Dine and N.~Seiberg,
  ``Exponential Hierarchy From Dynamical Supersymmetry Breaking,''
  Phys.\ Lett.\  B {\bf 140}, 59 (1984).
}

\lref\AffleckXZ{
  I.~Affleck, M.~Dine and N.~Seiberg,
  ``Dynamical Supersymmetry Breaking In Four-Dimensions And Its
  Phenomenological Implications,''
  Nucl.\ Phys.\  B {\bf 256}, 557 (1985).
}

\lref\LeeXH{
  T.~Lee and G.~H.~Wu,
  ``Nonlinear supersymmetric effective Lagrangian and goldstino  interactions
  at high energies,''
  Mod.\ Phys.\ Lett.\  A {\bf 13}, 2999 (1998)
  [arXiv:hep-ph/9811458].
}

\lref\HoKimTF{
  Q.~Ho-Kim and A.~Smailagic,
  ``Constrained Superfields,''
  Phys.\ Rev.\  D {\bf 29}, 2427 (1984).
}

\lref\ClarkAW{
  T.~E.~Clark and S.~T.~Love,
  ``Goldstino couplings to matter,''
  Phys.\ Rev.\  D {\bf 54}, 5723 (1996)
  [arXiv:hep-ph/9608243].
}

\lref\BrignoleME{
  A.~Brignole, F.~Feruglio, M.~L.~Mangano and F.~Zwirner,
  ``Signals of a superlight gravitino at hadron colliders when the other
  superparticles are heavy,''
  Nucl.\ Phys.\  B {\bf 526}, 136 (1998)
  [Erratum-ibid.\  B {\bf 582}, 759 (2000)]
  [arXiv:hep-ph/9801329].
}

\lref\KleinVU{
  M.~Klein,
  ``Couplings in pseudo-supersymmetry,''
  Phys.\ Rev.\  D {\bf 66}, 055009 (2002)
  [arXiv:hep-th/0205300].
}

\lref\BaggerGM{
  J.~A.~Bagger and A.~F.~Falk,
  ``Decoupling and Destabilizing in Spontaneously Broken Supersymmetry,''
  Phys.\ Rev.\  D {\bf 76}, 105026 (2007)
  [arXiv:0708.3364 [hep-ph]].
}

\lref\WeinbergKR{
  S.~Weinberg,
  ``The quantum theory of fields. Vol. 2: Modern applications,''
{\it  Cambridge, UK: Univ. Pr. (1996) 489 p}
}

\lref\LutyFK{
  M.~A.~Luty,
  ``Naive dimensional analysis and supersymmetry,''
  Phys.\ Rev.\  D {\bf 57}, 1531 (1998)
  [arXiv:hep-ph/9706235].
}

\lref\LiebrandPN{
  F.~Liebrand,
  ``Coset Based Formalism For The Nonlinear Realization of the Supersymmetry
  Algebra,''
  Phys.\ Rev.\  D {\bf 42}, 3457 (1990).
}

\lref\UematsuRJ{
  T.~Uematsu and C.~K.~Zachos,
  ``Structure Of Phenomenological Lagrangians For Broken Supersymmetry,''
  Nucl.\ Phys.\  B {\bf 201}, 250 (1982).
}

\lref\BrignoleCM{
  A.~Brignole, J.~A.~Casas, J.~R.~Espinosa and I.~Navarro,
  ``Low-scale supersymmetry breaking: Effective description, electroweak
  breaking and phenomenology,''
  Nucl.\ Phys.\  B {\bf 666}, 105 (2003)
  [arXiv:hep-ph/0301121].
}

\lref\DineII{
  M.~Dine, R.~Kitano, A.~Morisse and Y.~Shirman,
  ``Moduli decays and gravitinos,''
  Phys.\ Rev.\  D {\bf 73}, 123518 (2006)
  [arXiv:hep-ph/0604140].
}

\lref\AntoniadisUK{
  I.~Antoniadis and M.~Tuckmantel,
  ``Non-linear supersymmetry and intersecting D-branes,''
  Nucl.\ Phys.\  B {\bf 697}, 3 (2004)
  [arXiv:hep-th/0406010].
}

\lref\AntoniadisSE{
  I.~Antoniadis, M.~Tuckmantel and F.~Zwirner,
  ``Phenomenology of a leptonic goldstino and invisible Higgs boson decays,''
  Nucl.\ Phys.\  B {\bf 707}, 215 (2005)
  [arXiv:hep-ph/0410165].
}


\Title{
} {\vbox{\centerline{From Linear
SUSY to Constrained Superfields}
}}
\medskip

\centerline{\it Zohar Komargodski and Nathan Seiberg}
\bigskip
\centerline{School of Natural Sciences}
\centerline{Institute for Advanced Study}
\centerline{Einstein Drive, Princeton, NJ 08540}

\smallskip

\vglue .3cm

\bigskip
\noindent
We present a new formalism for finding the low-energy effective Lagrangian of Goldstinos and other fields.  This Lagrangian is written using standard superspace and the superfields are constrained to include only the light degrees of freedom.  The Goldstino resides in a (constrained) chiral superfield $X$ which is naturally identified at short distances.  This allows us to exactly compute the IR behavior of some correlation functions even in strongly coupled theories with SUSY breaking.  The Goldstino couplings above the scale of the matter superpartners are determined by identifying $X$ with the standard spurion.  At energies below the superpartners' scale, fermions, scalars (including Goldstone bosons) and gauge fields are also described by constrained superfields.  Our framework makes it easy to find the leading order terms in the Lagrangian and to control their corrections.  It simplifies the derivation of many known results and leads to new ones.

\Date{7/2009}

\newsec{Introduction}

One universal prediction of spontaneous breaking of rigid supersymmetry is a massless Weyl fermion, the Goldstino $G_\alpha$. In supergravity the Goldstino is eaten by the gravitino and becomes massive. This effect is negligible when the SUSY breaking scale $f$ is much smaller than $M_P^2$.  In that case the low-energy couplings of the gravitino are dominated by its Goldstino component and can be analyzed in the rigid limit.

Despite an extensive literature on Goldstino couplings, e.g~\refs{\VolkovIX\IvanovMY\IvanovMX\RocekNB\LindstromKQ\IvanovBPA
\UematsuRJ\SamuelUH\HoKimTF\LiebrandPN\ClarkAW\BrignoleFN
\LutyNP\BrignolePE\BrignoleSK\ClarkAA\BrignoleME\LeeXH\KleinVU\AntoniadisUK\AntoniadisSE
\DineII-\BaggerGM}, a simple and transparent presentation of the subject is still lacking.  In particular, many couplings of Goldstinos to Goldstinos and of Goldstino to matter particles are known.  However, some terms are not easily found using the known methods and it is not always clear how to identify the leading order terms and their corrections.  Also, some of the computations exhibit miraculous cancelations and lead to simple results whose origin is not obvious.

For example, if a certain theory breaks SUSY and has only the Goldstino at low energies the Lagrangian at very low energies should of course be
\eqn\basiclag{{\cal L}=-f^2+i\pa_\mu\bar G\bar\sigma^\mu G+\cdots~,}
where the ellipses stand for terms with more derivatives and more fermions. Some of these terms are described by the Akulov-Volkov (AV) Lagrangian \VolkovIX. But there are also higher order corrections which we would like to identify.

When a global symmetry is spontaneously broken the corresponding conserved charge does not exist because its correlation functions are IR divergent.  However, the conserved current and even commutators with the conserved charge do exist.  Therefore, although the one particle states are not in linear representations of the symmetry, the local operators are.  The same is true when supersymmetry is spontaneously broken in infinite volume.  (When supersymmetry is spontaneously broken in finite volume even the supercharge exists.)  The supercharge $Q_\alpha$ does not exist but the supercurrent and the (anti)commutators with $Q_\alpha$ exist.  This means that operators reside in supermultiplets  and we can construct superspace and superfields.

Here we will use this well known observation to give a new superspace description of Goldstinos and matter fields.  We will start in section 2 by reviewing the supersymmetry current and its supermultiplet.  This is the Ferrara-Zumino (FZ) multiplet $\CJ_{\alpha \alphadot}$ \FerraraPZ\ which satisfies the conservation equation
\eqn\FZint{\bar D^\alphadot \CJ_{\alpha \alphadot}= D_\alpha X~,}
where $X$ is a chiral superfield.  Studying a general theory with broken supersymmetry, we identify the IR limit of this $X$ as proportional to the Goldstino superfield.  Hence the microscopic operator $X$ includes both the order parameter for supersymmetry breaking and the Goldstino with a canonical normalization.  At low energies $X$ satisfies $X^2=0$.  This leads us to consider the nonlinear Goldstino superfield $X_{NL}$ as an ordinary chiral superfield constrained to satisfy
 \eqn\Goldcontaa{X_{NL}^2=0~.}
Its lowest component is proportional to $G^2$.  It is important that this description of the Goldstinos and their superfields allows us to compute the IR behavior of correlation functions of $X$ in any theory with broken supersymmetry.

In section 3 we write Lagrangians for $X_{NL}$.  We identify the AV Lagrangian as the leading Lagrangian with an appropriate scaling and discuss the higher order corrections.  We demonstrate our formalism in a particular example based on a simple theory of supersymmetry breaking with a single chiral superfield.

In section 4 we study couplings of the Goldstino to superfields.  This discussion is relevant at energies below $\sqrt f$ but above the masses of the superpartners. Here we identify $\Xnl$ with the known spurion, thereby establishing the existence of a spurion field in all SUSY-breaking models, including strongly coupled ones.

Section 5 describes the Goldstino couplings at very low energies, below the masses of the superpartners. We consider couplings to light fermions, light gauge fields and light scalars (including Goldstone bosons of global symmetries).  We embed these fields in superfields and impose constraints which remove the massive degrees of freedom.  The scalar from a chiral superfields $Q$, which we often refer to as a squark (but it could also be any other sfermion), is removed by imposing
 \eqn\QNLaa{X_{NL}Q_{NL}=0 ~.}
Fermions (e.g.\ Higgsinos) from chiral superfields $\CH$ are removed by imposing
 \eqn\HNLaa{X_{NL}\bar\CH_{NL} = {\rm chiral}~.}
If we want to leave only a real scalar out of a chiral superfield $\CA$ we impose
 \eqn\ANLaa{X_{NL}(\CA_{NL}- \bar\CA_{NL})=0 ~.}
Such fields can be used to describe Goldstone bosons of broken continuous symmetries or axions.
Under the broken symmetry, $\CA_{NL}$ transforms inhomogeneously as $\CA_{NL}\rightarrow \CA_{NL}+\alpha$
where $\alpha$ is a real transformation parameter.  Note that this transformation is consistent with \ANLaa.
We can also exponentiate $\CA_{NL}$ to find a superfield $\CR_{NL}= e^{i\CA_{NL}}$ with charge one. Then
\eqn\ANLaa{X_{NL}(\CR_{NL}\bar\CR_{NL}-1)=0 ~.}

Gauge fields are a bit more subtle.  It is well known that supersymmetric gauge theories are invariant under the gauge symmetry $V \to V + \Omega + \bar \Omega$ with chiral $\Omega$.  This symmetry is larger than that of non-supersymmetric theories. So if we want to describe a non-supersymmetric gauge theory using vector superfields, we need to restrict the gauge symmetry to that of the non-supersymmetric theory. Normally, this is achieved by choosing
the Wess-Zumino gauge.  In our framework it is the gauge choice
\eqn\VNLaa{X_{NL}V_{NL}=0~.}
The remaining gauge freedom $V_{NL} \to V_{NL} + \Omega_{NL} + \bar \Omega_{NL}$ is characterized by
\eqn\remga{X_{NL}\left(\Omega_{NL}+\bar\Omega_{NL}\right)=0~.}
Then, if charged matter fields are present and we embed them in superfields like $Q_{NL}$, $\CH_{NL}$ or $\CR_{NL}$ we can assign to them gauge transformation laws by multiplication by $e^{\Omega_{NL}}$.  In addition,
if the gauginos are massive we need to eliminate them.  This is achieved by imposing
\eqn\WNLaa{X_{NL}W_{NL\alpha }=0 ~.}

These constraints have an analogy in pion physics, although it is not usually discussed.
For instance, consider the situation of global symmetry breaking of $SO(3)$ to $SO(2)$. We can
describe the Goldstone bosons  by a ${\bf 3}$ vector $\vec n$ constrained to
satisfy $\vec{n} \cdot \vec n =1$.  This is analogous to our $\Xnl$ with the
constraint $\Xnl^2 =0$.  Then, using $\vec n$ we can describe other low energy excitations
by making them satisfy certain relations. For example, we can have a
three vector of fermions $\vec \psi$ in the ${\bf 3}$ of $SO(3)$.  Because of the
symmetry breaking, it breaks into a neutral fermion $\psi_0= \vec n \cdot \vec
\psi$ and two charged ones which can be described by the components of $\vec \psi$ subject to $\vec n \cdot \vec \psi =0$. In other words, $\vec n$ allows us to write $SO(3)$ invariant expressions
even when this symmetry is spontaneously broken.
Clearly, this is
analogous to our various constraints on the superfields~\Goldcontaa-\WNLaa.

In section 5 we also use our formalism to present explicit component terms involving Goldstinos and matter fields.  Here we easily reproduce many known results and derive new ones.

In several appendices we present some more technical details.

\newsec{The Supercurrent Multiplet and Spontaneous SUSY Breaking}

\subsec{The Supercurrent Multiplet}

We start this section with a review of the Ferrara-Zumino (FZ) multiplet~\FerraraPZ. It is a Lorentz vector, real superfield $\CJ_\mu$ which includes the supersymmetry current $S_{\mu\alpha}$, the energy-momentum tensor $T_{\mu\nu}$ and an R-current $j_\mu$. (The R-current is not necessarily conserved.) It satisfies the conservation equation\foot{For any vector $\ell_\mu$,
$$\ell_{\alpha\dot\alpha}=-2\sigma^\mu_{\alpha\dot\alpha}\ell_\mu~,\qquad \ell_\mu={1\over 4}\bar\sigma_\mu^{\dot\alpha\alpha}\ell_{\alpha\dot\alpha}~.$$
}
\eqn\defeq{\bar D^{\dot\alpha}\CJ_{\alpha\alphadot}=D_\alpha X~,}
with $X$ some chiral superfield.

Solving \defeq\ is straightforward
\eqn\finalresult{\eqalign{
&\CJ_\mu=j_\mu+\theta^\alpha\left(S_{\mu\alpha}+{1\over 3}(\sigma_\mu\bar\sigma^\rho
S_\rho)_\alpha\right)+\bar\theta_{\dot\alpha} \left(\bar S_\mu^{\dot\alpha}+{1\over
3}\epsilon^{\dot\alpha\dot\beta}(\bar S_\rho\bar\sigma^\rho\sigma_\mu)_{\dot\beta}
\right)\cr
&\qquad  +(\theta\sigma^\nu\bar\theta)\left(2T_{\nu\mu}-{2\over
3}\eta_{\nu\mu}T-{1\over4}\epsilon_{\nu\mu\rho\sigma} \pa^{[\rho}j^{\sigma]}\right) +{i\over 2}\theta^2\pa_\mu
\bar x-{i\over 2}\bar\theta^2\pa_\mu x +\cdots~\cr
&\pa^\mu T_{\mu\nu}=\pa^\mu S_{\mu\alpha}=0~,\qquad T_{\mu\nu}=T_{\nu\mu}~.}}
The ellipses represent terms with more $\theta$s.  They are determined by the lower order terms~\finalresult.
The last line represents the known conservation equations and the fact that the energy momentum tensor is symmetric.
In terms of these fields the chiral superfield $X$ is
\eqn\fermionrel{\eqalign{
&X=x(y)+\sqrt 2\theta\psi(y)+\theta^2 F(y)~,\cr
&\psi_\alpha={\sqrt2\over 3}\sigma_{\alpha\dot\alpha}^\mu\bar S_\mu^{\dot\alpha}~,\qquad F={2\over
3} T+i\pa_\mu j^\mu,}}
where we follow the conventions of~\WessCP\ with
\eqn\ydefa{y^\mu=x^\mu +i\theta\sigma^\mu\bar\theta ~.}

Using these expressions we can derive the SUSY variations of the supercurrent
\eqn\Qdaggertransf{\{\bar
Q_{\dot\beta},S_{\mu\alpha}\}=\sigma_{\alpha\dot\beta}^\nu\left(2T_{\mu\nu} -i\eta_{\nu\mu}\pa j+i\pa_\nu
j_\mu-{1\over 4}\epsilon_{\nu\mu\rho\sigma}\pa^{[\rho}j^{\sigma]}\right)~,}
\eqn\Qtransf{\{Q_{\beta},S_{\mu\alpha}\}=2i\epsilon_{\lambda\beta}\left( \sigma_{\mu\rho}\right)^{\lambda}_\alpha\pa^\rho \bar x~.}

We remark that the solution of \defeq\ is not unique. If a pair $(\CJ_\mu,X)$ satisfies the equation, so does the shifted pair
\eqn\newpair{\left(\CJ_\mu+i\pa_\mu(Y-\bar Y),X-{1\over 2}\bar D^2\bar Y\right)~.}
This transformation does not change the charges but it affects the improvement terms of the supercurrent and the energy momentum tensor.

In a superconformal theory $X=0$ and therefore we get \eqn\superconformal{T=0~,\qquad \pa_\mu j^\mu=0~,\qquad \bar\sigma^{\mu\dot\alpha\alpha} S_{\mu\alpha}=0~.}

Interestingly, the supercurrent multiplet requires the existence of an operator $x$. The need for this operator follows from \Qtransf, as a supersymmetry transform of the supercurrent (even though, as a Schwinger term, it does not affect the algebra of charges).
Thus, $x$ is a well defined operator in the UV, which can be expressed in terms of the elementary degrees of freedom.

As an example, a general sigma model
\eqn\sigmamodel{S=\int d^4\theta K(\Phi ^i, \bar\Phi^\ibar )
 + \int d^2 \theta
 W(\Phi^i) + \int d^2 \bar \theta \bar W(\bar \Phi^\ibar)~}
leads to
\eqn\genXs{\eqalign{
\CJ_{\alpha \alphadot} &=-{2\over 3} g_{i \ibar}(\bar
 D_\alphadot \bar \Phi^\ibar)( D_\alpha \Phi^i)
 +i{2\over 3}\left(\partial_i K
 \partial_{\alpha\alphadot} \Phi^i - \partial_\ibar K
 \partial_{\alpha\alphadot} \bar \Phi^\ibar \right)  + i
 \partial_{\alpha\alphadot}( Y(\Phi) - \bar Y( \bar\Phi)) \cr
 &= 2g_{i \ibar}( D_\alpha  \Phi^i)(\bar D_\alphadot \bar \Phi^\ibar)
 - {2\over 3}[D_\alpha, \bar D_\alphadot ] K  + i
 \partial_{\alpha\alphadot}( Y(\Phi) - \bar Y( \bar\Phi))~, \cr
 X & =4 W-{1\over3}\bar D^2K~ - {1\over 2}\bar D^2 Y (\Phi)~.}}
This expression has already been given in \ClarkBG. Note that we have included
an arbitrary holomorphic function $Y(\Phi)$ corresponding to the ambiguity \newpair.  Here it is identified with the freedom in K\"ahler transformations
 \eqn\Kahlert{K \to K+{3 \over 2} (Y + \bar Y)~.}
This ambiguity plays a central role in \KomargodskiPC.

\subsec{SUSY Breaking}

As we emphasized in the introduction, the multiplets $\CJ_{\alpha\alphadot}$ and $X$ can be used even when SUSY is spontaneously broken.  These operators are well defined and can be followed consistently along the RG flow down to the deep IR.  Let us demonstrate it in a trivial example -- a free theory which exhibits SUSY breaking.  Its Lagrangian is
\eqn\lagonei{\CL=\bar\Phi \Phi\bigr|_{\theta^2\bar\theta^2}+\left(f\Phi\bigr|_{\theta^2}+c.c.\right)~.}
The chiral superfield
\eqn\defchiral{\Phi=\phi(y)+\sqrt2\theta\psif(y)+\theta^2F_\phi(y)}
includes the free fermion $({\psif})_\alpha$ which we identify as the Goldstino $G_\alpha$.  Using \genXs\
\eqn\Xfree{\eqalign{
&\CJ_{\alpha\dot\alpha}=i{2\over 3} \left(\bar\Phi\pa_{\alpha\dot\alpha}\Phi-\Phi\pa_{\alpha\dot\alpha}\bar\Phi\right)- {2\over 3}\bar D_{\dot\alpha}\bar\Phi D_\alpha\Phi~, \cr
&X={8\over 3}f\Phi~.}}
Therefore, we see that the operator $X$ is proportional to the superfield $\Phi$.

More generally, standard arguments about symmetry breaking show that when supersymmetry is spontaneously broken, the low-energy supercurrent is expressed in terms of the massless Goldstino $G_\alpha$ as
\eqn\suplow{S_{\mu\alpha}= \sqrt2f\sigma_{\mu\alpha\alphadot} \bar G^\alphadot+f'(\sigma_{\mu\nu})^\beta_\alpha\pa^{\nu}G_\beta~ + \cdots~,}
with some constants $f$ and $f'$.
Note that the term with $f'$ is an improvement term (associated with the ambiguity~\newpair) and therefore can be ignored.  We see that the spin $(1,{1\over 2})$ component of the supercurrent essentially decouples from the low-energy physics, while the $(0,{1\over 2})$ representation is proportional to the Goldstino field. We conclude that as in the trivial example \Xfree, at long distances the $\theta$ component of $X$, which is the spin ${1\over 2}$ projection of the supercurrent~\fermionrel, is the Goldstino
\eqn\Xtheta{X_{NL}\bigr|_{\theta^\alpha}= \sqrt 2 G_\alpha~,}
where, for reasons which will soon be obvious, we denote the long distance expression of ${3  \over 8 f}X$ as $X_{NL}$.

In the trivial example \Xfree\ the lowest component $x$ of the superfield $X$ is proportional to the free scalar $\phi$.  In general, the low-energy Goldstino is not accompanied by a massless scalar.  Then, the bosonic operator $x$ cannot create a one particle state.  Instead, the simplest bosonic state it can create is constructed out of two Goldstinos.  When supersymmetry is broken in finite volume the states in the Hilbert space are in supersymmetry multiplets.  Then, the supersymmetric partner of a one Goldstino state $Q^\dagger_\alphadot |0\rangle$ is a ``two Goldstino state'' $Q^\dagger _{\dot 1}Q^\dagger_{\dot 2}|0\rangle$.  At infinite volume the supercharge does not exist and the zero momentum Goldstino state is not normalizable.  However, the finite volume intuition is still valid.  The operator $\psi_\phi$ creates a one Goldstino state and its superpartner $\phi \sim x$ creates a two Goldstino state.

More explicitly, denoting the first component of $X_{NL}$ by $x_{NL}$, the supersymmetry transformations are
 \eqn\XnlSUS{\eqalign{
 &[\xi Q,x_{NL}]= \sqrt 2 \xi G \qquad , \qquad [\bar \xi \ \bar Q,x_{NL}]=0~,\cr
 &[\xi Q,G_\alpha]= \sqrt 2 \xi_{\alpha} F \qquad , \qquad [\bar \xi \ \bar Q,G_\alpha ]= i \sqrt 2 \sigma^\mu_{\alpha \alphadot} \bar \xi^\alphadot\partial_\mu x_{NL}~, \cr
 &[\xi Q,F]=0   \qquad \quad \qquad , \qquad [\bar \xi \ \bar Q ,F]= i \sqrt 2 \bar \xi_\alphadot \bar \sigma^{\mu \alphadot\alpha} \partial_\mu G_\alpha ~.}}
This fixes $x_{NL}= {G^2 \over 2 F}$ up to an uninteresting additive constant. We conclude that
\eqn\solutionX{X_{NL}={G^2\over 2F}+\sqrt2\theta G+\theta^2 F ~,}
where all the fields are functions of $y^\mu$ \ydefa.  Clearly, \solutionX\ satisfies the interesting operator identity
\eqn\niliden{X_{NL}^2=0~.}

The expression for $X_{NL}$ can also be written as
\eqn\Xnlal{X_{NL}= F \Theta^2 ~, \qquad \Theta =\theta  + {G \over \sqrt 2F}~,}
where again, all the fields are functions of $y^\mu$.  This makes it obvious that $X_{NL}^2 = 0$.  The combination $ \Theta=\theta  + {G \over \sqrt 2F}$ is reminiscent of expressions found in the literature on nonlinear realizations (for a summary of the various approaches see~\SamuelUH), except that we have an independent auxiliary field $F$ rather than its vev $f$.

We conclude that the Goldstino always resides in a chiral superfield $\Xnl$ which satisfies $\Xnl^2=0$.  Furthermore, this chiral superfield is the IR limit of the microscopic superfield $X$.  Surprisingly, this result is true both for $F$-term and for $D$-term breaking because it relies only on the existence of the chiral operator $X$.\foot{The only exception is the situation with pure $D$-term breaking which occurs when a {\it tree-level} FI-term is present for an unbroken $U(1)$ gauge theory.  In this case the operator $X$ is not gauge invariant~\KomargodskiPC\ and we do not discuss it here.  However, our discussion does apply in the Higgs phase of the FI-model in which the low energy limit of $X$ is gauge invariant.}  Thus, our discussion is applicable in all the models of dynamical SUSY breaking.

Several authors have also included the Goldstino field in superfields.  Samuel and Wess~\SamuelUH\ studied various superfields which are not chiral (they are roughly $D_\alpha X_{NL}$).  Rocek considered a chiral superfield similar to our $X_{NL}$ except that it satisfies another equation $X_{NL} \bar D^2 \bar X_{NL} =4 f X_{NL}$ \RocekNB.  This equation eliminates the $\theta^2$ component of $X_{NL}$ which we prefer not to do.  We will later see that in some cases this equation arises as an equation of motion.  Finally Lindstrom and Rocek described the Goldstino in a vector superfield~\LindstromKQ.

It is instructive to compare the situation here with the theory of ordinary Goldstone bosons like pions.  Clearly, the decay constant $f$ here is analogous to the decay constant $f_\pi$ of pion physics.  Both of them are well defined.  However, in pion physics the order parameter $\langle \psi\psi\rangle$ for chiral symmetry breaking does not have a universal definition with a precise normalization (it suffers from wave function normalization).  In our case the order parameter for supersymmetry breaking is the energy momentum tensor which resides in the same multiplet as the supersymmetry current and hence it has a well defined normalization.  Therefore, our $X$ is completely well defined. Correspondingly, the operator $\psi\psi$ of pion physics acts as an interpolating field for pions, but its normalization is not meaningful.  In our case $X$ is used both as an order parameter for supersymmetry breaking and as an interpolating field for Goldstinos with well defined normalization.

The analogy with pion physics also clarifies the meaning of our constraint \niliden.  It arises from removing the massless scalar in $X$.  This is analogous to describing pion physics by starting with a linear sigma model with a sigma field.  Removing the sigma field is implemented by imposing a constraint of the field $UU^\dagger =1$ which is analogous to our $X_{NL}^2=0$.

Finally, we would like to comment about another application of our identification of $X$ in the low-energy theory.  In every microscopic theory we can identify $X$ in the ultraviolet using \defeq.  Then \solutionX\ describes its low-energy limit. We can therefore calculate, for example, correlation functions of $x$ at large separations even in strongly coupled models which break supersymmetry. We see that also ``incalculable" models of SUSY breaking (like the $SU(5)$ theory of \AffleckVC\ or the $SO(10)$ theory of \AffleckMF) have a solvable sector at long distances. The operator $x$ interpolates between the vacuum and a state with two Goldstinos, with a universal normalization. This is because at long distances the leading behavior of $x$ is proportional to that of $G^2$.  Therefore,
\eqn\schcorr{\lim_{|{\bf r}|\rightarrow\infty}\langle x({\bf r})\bar x(0)\rangle= \left({4 \over 3 \pi^2}\right)^2 {1\over |{\bf r}|^6}~,}
which is independent of the details of the microscopic theory and its coupling constants.  In a similar fashion we can calculate the long distance limit of any correlation function of $x$ and $\bar x$.

\newsec{Lagrangians for $X_{NL}$}

\subsec{General Analysis}

Now we will use $X_{NL}$ to write supersymmetric effective Lagrangians.  We start without including derivatives. Then the most general supersymmetric Lagrangian (without derivatives) subject to the constraint $X_{NL}^2=0$ is
\eqn\AVaction{{\cal L}=\int d^4\theta X_{NL}\bar X_{NL} +\int d^2\theta fX_{NL}+\int d^2\bar \theta   f  \bar X_{NL}~,}
where without loss of generality we take $f$ to be real.  This looks like the trivial example~\lagonei\ except that the superfield is constrained.  This difference removes the massless scalar field and introduces nonlinearities.

More explicitly, the constraint can be solved as \solutionX
 \eqn\solvcon{X_{NL}={G^2\over 2F}+\sqrt2\theta G+\theta^2 F~.}
Substituting this in \AVaction, we derive the component Lagrangian
\eqn\AVlag{{\cal L}=i\partial_\mu\bar G\bar\sigma^\mu G +\bar F F+{\bar G^2\over 2\bar F}\partial^2\left({G^2\over 2F}\right)+\left(fF +c.c.\right)~.}
The equations of motion of the auxiliary fields $F,\bar F$ are
\eqn\auxi{\eqalign{
&F + f-{\bar G^2\over 2\bar F^2}\partial^2\left({G^2\over 2F}\right)=0~, \cr
&\bar F+f-{G^2\over 2 F^2}\partial^2\left({\bar G ^2\over 2\bar F}\right)=0~.}}
They are solved by
\eqn\exactsol{\eqalign{
&F=- f\left(1+{\bar G ^2\over 4f^4}\partial^2{ G ^2}-{3\over 16f^8} G ^2\bar G ^2\partial^2 G ^2\partial^2\bar G ^2\right)~, \cr
&\bar F=-f\left(1+{ G ^2\over 4 f^4}\partial^2{\bar G ^2}-{3\over 16f^8} G ^2\bar G ^2\partial^2 G ^2\partial^2\bar G ^2\right)~.}}
Upon substituting this solution back in \AVlag\ we find
\eqn\Goldstinolag{{\cal L}_{AV}=-f^2+i\partial_\mu\bar G \bar\sigma^\mu G +{1\over 4f^2}\bar G ^2\partial^2 G ^2-{1\over 16f^6} G ^2\bar G ^2\partial^2 G ^2\partial^2\bar G ^2 ~.}
This is equivalent to the Akulov-Volkov Lagrangian \VolkovIX.  We conclude that \AVaction\ is a simple description of the AV Lagrangian.

What is the equation of motion which follows from \AVaction?  To answer this question we add a superpotential term ${1\over 2}\lambda X_{NL}^2$ to \AVaction\ in order to implement the constraint where $\lambda$ is a Lagrange multiplier chiral superfield.  Now the equations of motion are $X_{NL}^2=0$ and ${1 \over 4} \bar D^2 \bar X_{NL}  = \lambda X_{NL} + f$.  The first is our constraint and the second determines $G$ and $F$.  However, if we multiply the second equation by $X_{NL}$ and use the constraint, we derive
 \eqn\rocekeq{X_{NL} \bar D^2 \bar X_{NL}=  4 f X_{NL}~.}
This equation contains less information than the full equation of motion.  It determines $F$ but does not determine $G$.  One way to think about it, is as the variation equation of the Lagrangian~\AVaction\ under $\delta \Xnl= \chi\Xnl$ with an infinitesimal chiral superfield $\chi$.

Rocek used \rocekeq\ to constrain his Goldstino superfield~\RocekNB.  We see that our formalism leads to this equation as a consequence of the equation of motion.  However, it is important to stress that once we include corrections to~\AVaction, or terms with additional fields (see below), the equations of motion are modified and~\rocekeq\ no longer holds.  Therefore, for the minimal theory our on shell description coincides with that of~\RocekNB.  When various corrections are added both descriptions are valid but they differ by field redefinitions.

It is again instructive to compare this discussion with the theory of pions.  The theory \AVaction\ appears free but the constraint makes it nonlinear.  This is similar to the leading order Lagrangian for pions which is not free only because the field $U$ is constrained to be unitary.

Both theories \AVaction\ and the leading order pion Lagrangian are subject to higher derivative corrections.  In the case of pions the leading order term has two derivatives and the corrections are suppressed by additional derivatives divided by powers of some high scale, e.g.\ $f_\pi$.  Clearly, this cannot be so simple in our case.  The Lagrangian \Goldstinolag\ includes terms with different numbers of derivatives and various factors of $f$. The reason for this is as follows.  Since pions represent the breaking of a global symmetry, it is natural to assign them the same dimension as the corresponding transformation parameter; i.e.\ dimension zero.  In our case the transformation parameter of supersymmetry has dimension $-{1\over 2}$.  Hence it is more natural to assign scaling $S=-{1\over 2}$ to $G$, which differs from its canonical dimension.  Equivalently, in the case of pions the natural field is the dimensionless field $\pi/f_\pi$, and in our case the natural field is the dimension $-{1\over 2}$ field $G/f$.  In terms of superfields, this means that we should assign $S(X_{NL})=-1$. Consequently, a term with $k$ derivatives and $l$ fermions, has scaling $S=k-{l\over 2}$.  The leading order Lagrangian \AVaction\ or \Goldstinolag\ has $S=0$. Its corrections which have more derivatives have higher values of $S$. A related scaling assignment was discussed in~\refs{\LutyFK,\LutyNP}.

We should make an important comment about the possible corrections to the leading order Lagrangian~\AVaction.  Obviously, we can restrict the various correction terms by using integration by parts and the constraint $\Xnl^2=0$. For example, this allows us to absorb $\int d^2 \theta \Xnl \bar D^2 \bar X_{NL}$  in the normalization of $\Xnl$ and the value of $f$.   A bit less trivial is the freedom in performing field redefinitions.  We can redefine $\Xnl$ while preserving the constraint e.g.\ $\Xnl \to \Xnl + \chi\Xnl $ where $\chi$ is a chiral superfield.\foot{It is important to stress that such a redefinition does not affect our identification of $\Xnl$ as the IR limit of the operator $X$ (normalizing $\Xnl$ such that it has a canonical kinetic term, the redefinition $\delta \Xnl= \chi\Xnl$ is always negligible at long distances).}  As an example, this allows us to remove $S=0$ terms like
\eqn\redop{\int d^4\theta \bar X_{NL} \Xnl \bar D^2 \bar\Xnl~.}
The equation \rocekeq\ allows us to simplify the analysis of such field redefinitions. For example, using \rocekeq\ the operator \redop\ is easily seen to be redundant.  In fact, it is easy to show that \AVaction\ are the only $S=0$ terms. The proof of this fact is presented in appendix A.

Higher order corrections terms like the $S=2$ term
\eqn\highSterm{\int d^4 \theta |\partial \Xnl|^2}
are suppressed by additional powers of higher scales.  These could be either $\sqrt f$ or perhaps other scales.  In the next subsection we will consider an explicit example where the operator \highSterm\ is suppressed by powers of a scale which is lower than $\sqrt f$.  It is important that such high order corrections lead to terms with either more derivatives or more fermions than the leading order Lagrangian \AVaction.

A curious property of the AV Lagrangian is that it has an accidental R-symmetry under which $X_{NL}$ has charge 2. It must be there because the AV Lagrangian is universal and there are known examples of R-invariant theories which break SUSY. However, higher scaling corrections can break this symmetry. For instance the $S=2$ term
\eqn\Rbreaking{\int d^4 \theta \ \Xnl D^2(\Xnl)\partial^2\Xnl~,}
or equivalently
 \eqn\Rbreakinge{\int d^2 \theta \ \Xnl (\partial^2\Xnl)^2}
breaks the R-symmetry by 4 units.\foot{It is easy to show that there is no $S=2$ operator which violates the R-symmetry by two units.  However, it is possible to completely break the R-symmetry (more precisely, break to a ${\bf Z}_2$ which is in the Lorentz group) using higher scaling operators. } The component expression with the least number of fermions and derivatives which arises from \Rbreaking\Rbreakinge\ is of the form $\left(\partial^2(G^2)\right)^2$.

\subsec{Example}

In this subsection we analyze a simple model of supersymmetry breaking which demonstrates our general discussion above.  Unlike the free theory \lagonei, here there is no massless scalar and hence we will see how the constraint arises.  In order to remove the massless scalar we modify the K\"ahler potential of \lagonei:
 \eqn\efflag{\eqalign{
 &K= \Phi\bar \Phi- {c\over M^2} \Phi^2 \bar\Phi^2-\left({\tilde c\over M^2} \Phi^3\bar\Phi+c.c.\right)~, \cr
 &W_{eff}=f\Phi ~. \cr
 }}
Here $c, \tilde c$ are dimensionless numbers of order one.  We choose the sign of the real constant $c$ to be positive.  Without loss of generality we can take $\tilde c$ and $f$ to be real and non-negative.  Such a theory can arise as the low-energy Lagrangian below some scale $M$ after neglecting higher order terms in $1 \over M$.  It is valid for
 \eqn\energyE{\sqrt f \ll E \ll M~.}
For $\tilde c=0$ the theory has an R-symmetry which is explicitly broken for nonzero $\tilde c$.

For example, the low-energy limit of the standard O'Raifeartaigh model looks like~\efflag.  Since the O'Raifeartaigh model preserves an R-symmetry, it has $\tilde c=0$.  The term with $c$ arises at one loop.  Other UV theories without an R-symmetry can lead to non-zero $\tilde c$.

The scalar potential of this theory is
 \eqn\scalarpot{V={f^2 \over K_{\Phi,\bar\Phi}}= f^2 \left[ 1 + {4 c\over M^2} \phi \bar\phi+{3 \tilde c\over M^2} \phi^2+{3  \tilde c\over M^2} \bar\phi^2+\CO\left({1\over M^3}\right) \right]~,}
whose minimum is at $\phi=0$.  The two bosons have masses
 \eqn\massespm{m_\pm^2={f^2 \over M^2} (4c \pm 6 \tilde c)~.}
Hence the vacuum at $\phi=0$ is stable only for
 \eqn\crange{0\le \tilde c \le {2 \over 3} c}
(this is the reason we chose $c$ to be positive).  For $\tilde c={2 \over 3}c$ there is a massless particle and higher order corrections are important. The latter case can be obtained from UV completions with spontaneously broken R-symmetry (e.g.\ the  3-2 model~\AffleckXZ\ or the example in Appendix E). For simplicity we concentrate on the region
 \eqn\crange{0\le \tilde c < {2 \over 3} c~.}

Let us integrate out the massive bosons.  Using
\eqn\Lagra{\eqalign{
{\cal L}=&-f^2+|F_\phi+f|^2-{c\over M^2}\left|2\phi F_\phi-\psi^2\right|^2-3{\tilde c\over M^2}\left(\left(\phi^2F_\phi-\phi\psi^2\right)\bar F_\phi+c.c.\right) \cr
& + {\rm terms\ with\ derivatives }}}
the zero momentum equation of motion of $\phi$ is
\eqn\eom{2c\left(2\phi F_\phi-\psi^2\right)\bar F_\phi+3\tilde c\left(2\bar\phi\  \bar F_\phi-\bar\psi^2\right)F_\phi=0~.}
Its solution
\eqn\sol{\phi={\psi^2\over 2F_\phi}}
is independent of $c, \tilde c$. In other words, it is independent of the details of the high-energy physics.

We see that upon substituting \sol\ back into $\Phi$ we discover that in the IR $\Phi^2 =0$. Hence, it satisfies the same constraint as $X_{NL}$. To make the relation more precise, we calculate the operator $X$ defined in~\defeq\ in the UV model \efflag.  Using \genXs\ and the equation of motion we find in the UV
\eqn\XintFterm{X={1\over3}\left(8f\Phi-{c\over M^2}\Phi^2\bar D^2(\bar \Phi)^2 - 8 f {\tilde c \over M^2} \Phi^3+\CO\left({1\over M^4}\right)\right)~.}
As we have seen in our discussion above, well below the mass of the scalar field $\phi$ the superfield $\Phi$ satisfies $\Phi^2=0$. Due to this property we get at low energies ($E\ll {f\over M}, M$)
\eqn\XintFtermi{X={8\over3}f\Phi~,}
and hence in the IR we identify
 \eqn\PhiXNLid{\Phi=X_{NL}~.}

Finally, we compute the low-energy effective theory.  We substitute the zero momentum solution \sol\ back into the microscopic Lagrangian \efflag\ and express it in terms of $X_{NL}$.  Clearly, due to the property $X_{NL}^2=0$ the quartic terms in the microscopic theory do not contribute at leading order and the only contribution arises from the kinetic term; i.e.\ we end up with the Lagrangian \AVaction.

Let us discuss the corrections to this effective Lagrangian.  Clearly, its cutoff is $m_\pm \sim {f\over M}$ -- the energy scale at which the additional scalar shows up. Therefore, the Lagrangian has higher derivative corrections which are multiplied by powers of $M\over f$.  This is in addition to various other corrections we neglected at our starting point \efflag\ which are multiplied by $1\over M$.  These two kinds of corrections have higher scaling $S$.  Therefore, our truncation of the effective theory to \AVaction\ is meaningful.  Furthermore, since for nonzero $\tilde c$ our example does not have an R-symmetry, we expect higher scaling operators like~\Rbreaking\ to be present.

\newsec{Coupling Goldstinos to Superfields}

In this section we couple Goldstinos to a theory like the MSSM.  We have in mind a theory which is characterized by some renormalizable supersymmetric couplings as well as a set of soft supersymmetry breaking terms.  For simplicity we take all the dimensionful supersymmetric terms (like $\mu$) and all the soft terms to be of the same order of magnitude denoted $m_{soft}$.  Furthermore, we assume that the scale $\mu \sim m_{soft}$ is separated from a larger scale $\Lambda$ where additional physics might be present:
\eqn\MSSMa{\mu \sim m_{soft} \ll \Lambda ~.}
This allows us to view the MSSM as an effective theory at energies
\eqn\rangeE{\mu \sim m_{soft} \ll E \ll \Lambda ~,}
which is cut off at the scale $\Lambda$.

In this case it is common to write the soft breaking terms using a dimensionless spurion field \GirardelloWZ\ $Y= \theta^2 m_{soft}$. Then supersymmetric or non-supersymmetric corrections to the MSSM are suppressed by $1\over \Lambda^{d-4}$ where $d$ is known as the ``effective dimension'' (see e.g.~\DineXI).

We would like to replace the explicit supersymmetry breaking in the MSSM by spontaneous breaking with some $f$.  A complete microscopic theory involves adding degrees of freedom at the scale $\sqrt f$ which, because of our assumption \MSSMa, must satisfy $\Lambda \roughly< \sqrt f$.  We will not do that here, nor will we explain the hierarchy  $m_{soft} \ll \sqrt f$ (this can arise, as in gauge mediation, from loop factors or couplings through heavier messengers).  Instead, we will add only the Goldstino field $G_\alpha$, or more precisely the Goldstino superfield $\Xnl$.

The coupling of this theory to Goldstinos is easily found as follows.  Since $\langle X_{NL} \rangle = \theta^2 f$, it is natural to replace the spurion in the Lagrangian by
 \eqn\YrepX{Y \to {m_{soft}\over f} X_{NL}~.}
Then we add to the supersymmetric terms and the Lagrangian for $\Xnl$ \AVaction\ the following couplings.  First, there are couplings to chiral superfields $Q^i$
\eqn\lagchi{\CL_{soft}=- \int d^4\theta \left|{\Xnl\over f}\right|^2(m^2)_{i}^j(Q e^{ V}\bar Q )^{i}_j+\int d^2\theta {\Xnl\over f}\left( -{1\over 2}B_{ij} Q^i Q^j+{1\over 6}A_{ijk} Q^i Q^j Q^k \right) +c.c.~,}
where $m$, $B$ and $A$ are the standard soft masses and trilinear couplings.  Using spurions, the $A$ and $B$ terms can also be written as $D$-terms -- an integral over all of superspace.  In this case the resulting terms after the substitution \YrepX\ appear to be different than the terms in \lagchi.  However, a redefinition of $\Xnl$ by a shift proportional to $\chi\Xnl $ with an appropriate chiral superfield $\chi$ brings them to the form \lagchi\ plus higher dimension operators which we suppress.  Second, there are couplings to vector multiplets
\eqn\lagvec{\CL_{soft}=\int d^4\theta \left|{\Xnl\over f}\right|^2\xi D^\alpha W_\alpha+\int d^2\theta {\Xnl \over f} m_\lambda W_\alpha W^\alpha+c.c.~,}
where $\xi$ is a FI-term and $m_\lambda$ is the gaugino mass.  The FI-term is not usually written because it can be removed by a change of variables
\eqn\change{V'=V-g^2\xi\left|{\Xnl\over f}\right|^2~,}
which, due to $X_{NL}^2=0$, just changes the coefficients in \lagchi.

Normally, the use of spurions involves an assumption that the breaking of supersymmetry can be parameterized by a chiral spurion $Y$.  Our discussion here which starts with the operator equation \FZint\ and the chiral superfield $X$ shows that such a $Y$ always exists.  It exists even if the theory which spontaneously breaks supersymmetry is strongly coupled and does not have a simple description in terms of weakly coupled chiral superfields which exhibit tree level supersymmetry breaking.

Clearly, this nonrenormalizable Lagrangian leaves out physics at the scale $\Lambda$ (e.g.\ $\sqrt f$).  The Goldstino couplings in this Lagrangian involve high dimension operators and hence their loops have to be cut off.  Such loops were recently considered (using another formalism for the Goldstinos) in~\BaggerGM.

An important question is the identification of the corrections to this Lagrangian.  In the pure Goldstino problem we have identified the scaling $S$ which assigns $S(\Xnl)=-1$ and that helped us organize the corrections to the leading order AV Lagrangian \AVaction.  Here, we identify the field $\Xnl$ with the dimensionless spurion $Y$ \YrepX.  Let us see what this tells us about the corrections to the Lagrangian.

We assign the $S$ value of the chiral and vector superfields to be the same as their canonical dimensions.  Then the lowest scaling terms in the Lagrangian are the leading order Goldstino terms \AVaction\ which have $S=0$.  In section 3 we discussed some of the higher $S$ terms involving only Goldstinos.  The leading order terms with matter and gauge fields are the couplings \lagchi\lagvec\ which have $S=2,3$.  The supersymmetric terms involving matter and gauge fields have $S= 3, 4$.  Then, there are various couplings of these fields to $\Xnl$ with larger $S$, e.g.\ the $S=4$ term
\eqn\sameScou{\int d^4 \theta \ | \Xnl|^2  |\partial Q |^2 ~.}
Fortunately, our assumption \MSSMa\ means that such terms do not contribute to low momentum processes.  The reason for that is that even though they have the same $S$ as the supersymmetric terms, they have higher $S$ than the leading order couplings of Goldstinos and matter fields \lagchi\lagvec.  Hence, they lead to terms with more Goldstinos or more derivatives than \lagchi\lagvec.  More explicitly, terms like \sameScou\ arise from the spurion coupling $\int d^4 \theta \ |Y |^2  |\partial Q |^2 $  using the substitution \YrepX.  According to the standard dimension assignment this term has effective dimension 6; i.e.\ it is suppressed by $1 \over \Lambda^2$.  It reflects higher energy physics which is beyond the MSSM.  This means that the coefficient of \sameScou\ is of order $m_{soft}^2\over f^2\Lambda^2$.  In components it leads to several interactions including the coupling of 2 Goldstinos 2 matter fermions and 2 derivatives with coefficient $m_{soft}^2\over f^2 \Lambda^2$.  Evaluating the same amplitude using the couplings in \lagchi\lagvec\ including superpartner  intermediate lines we find a larger effect of order $1\over f^2 $.  We will discuss this in more detail below.

A complementary way to see that terms like \sameScou\ can be ignored is to use a different scaling counting. Previously we assigned $S(X_{NL})=-1$.  Instead, motivated by the standard dimension of the spurion $Y$, we can define $\tilde S(\Xnl)=0$. This means that the $F$-term carries $\tilde S(F)=1$, which is physically desirable since the limit $m_{soft}\ll\sqrt f$ suggests that indeed the vacuum energy is {\it relevant}. With this scaling,~\sameScou\ clearly has higher scaling than all the other terms, and therefore it should be suppressed by positive powers of $m_{soft}/\sqrt f$ and consequently dropped.

Let us summarize this section by recapitulating the different roles played by the operator $X$:
\item{1.} In the UV description of the theory it is an operator on the right hand side of $\bar D^\alphadot\CJ_{\alpha\alphadot}$~\FZint.  Here it is a measure of the violation of conformal invariance.
\item{2.} The expectation value of its $\theta^2 $ component is the order parameter for supersymmetry breaking.
\item{3.} When supersymmetry is spontaneously broken $X$ flows at low energies  to $X_{NL}= {3 \over 8f} X$ from \solutionX\ which is the Goldstino superfield.
\item{4.} As in \YrepX, $X$ generalizes the usual notion of a spurion.

\newsec{Matter Couplings}

In the previous section we have seen that the leading terms describing the interactions of the Goldstino particle with the MSSM superfields depend only on the soft parameters and the SUSY breaking scale $f$. These interactions allow us to extract the interesting information regarding physics at or above the soft breaking scale $m_{soft}$.

Our goal in this section is to develop tools for finding the interactions at energies below the soft scale
\eqn\enebelow{E \ll m_{soft}~.}
One way to derive these interactions is to start with the MSSM Lagrangian including the Goldstino couplings \lagchi\lagvec\ and to integrate out the superpartners.  Here we will follow another procedure which  directly determines these interactions without relying on a more microscopic description like the MSSM.

Our description has several advantages:
\item{1.}  Starting with the MSSM and the Goldstino couplings \lagchi\lagvec, as in section 4, we assumed \MSSMa\ and in particular $m_{soft} \ll \sqrt f$.  Then we could neglect high effective dimension terms in the MSSM as well as terms like \sameScou\ which depend on higher energy physics.  A formalism which is intrinsic to the low energy particles \enebelow\ has a larger range of validity because it allows $m_{soft} \roughly< \sqrt f $.
\item{2.}  Many authors have constructed a low energy effective Lagrangian in the range \enebelow\ by integrating out the superpartners \refs{\BrignoleFN,\BrignolePE,\BrignoleSK,\LeeXH,\BrignoleCM}.  These computations exhibit ``miraculous cancelations'' and lead to relatively simple final answers.  In another approach~\LutyNP\ one has to go through a nonlinear change of variables and calculate the results order by order in the number of Goldstinos and then integrate out the superpartners.  Again, this leads to simple answers.  Our formalism will make these ``miraculous cancelations'' and simplifications manifest.  It will explain the reason behind the simplicity of the final expressions in many cases and will allow us to re-derive and significantly extend many of the results in the literature.

Some authors, e.g.\ \refs{\ClarkAW,\ClarkAA,\ClarkBG} have focused on the light degrees of freedom and have used nonlinear realizations of supersymmetry to find their Lagrangian.\foot{In addition, an analysis in the context of string theoretic models was undertaken in~\AntoniadisUK, with an accompanying phenomenological study in~\AntoniadisSE.}  Although, this is technically correct, it turns out to be challenging to use this approach to find the most general Lagrangian and to control the corrections to the leading order terms. We will see some examples of this below.

Our basic idea is similar to the approach to the Lagrangian of the Goldstino around~\efflag; we write the Lagrangian in superspace using various constrained superfields. This makes supersymmetry manifest and quickly allows us to see that certain terms vanish without complicated calculations. In addition, it sheds light on the origin of various corrections.

Some of the terms in the effective action originate from the  kinetic terms of the light fields.  Their coefficients are universal -- they depend only on the SUSY breaking scale $f$.

Other terms have the same or higher scaling and their coefficients are not universal.  If at higher energies a description like the one in section 4 with the assumption \MSSMa\ is valid, then some of the non-universal terms are calculable from the MSSM data.  Other higher order terms are not easily determined and their coefficients depend on the details of the high energy theory.

\subsec{Fermions}

In this subsection we couple Goldstinos to light fermions. We will embed the fermions in chiral superfields.  Here our presentation will not follow the route we took in studying Goldstinos in sections 2 and 3. Instead, we will start with a microscopic example.

We study a system of two chiral superfields $\Phi$ and $Q$ interacting through
\eqn\toymatter{\eqalign{
&{ K}=\Phi \bar\Phi + Q \bar Q-{c\over M^2}\Phi^2\bar\Phi^2-{\hat c\over M^2}Q\bar Q\Phi\bar \Phi~, \cr
&{ W}= f\Phi~.}}
Here $M$ is some higher energy scale which is also the UV cutoff of this nonrenormalizable theory and $c$ and $\hat c$ are dimensionless coefficients.  We will refer to the generic matter field $Q$ as a quark (although it can be any other fermion).

 As in section 3, we want to integrate out the scalar components of $\Phi$ and $Q$.  The Lagrangian is
\eqn\Lagra{\eqalign{&{\cal L}=-f^2+|F_\phi+ f|^2+F_q\bar F_q-{c\over M^2}\left|2\phi F_\phi-\psi_\phi^2\right|^2  - {\hat c\over M^2}\left|q F_\phi+\phi F_q-\psi_q\psi_\phi\right|^2\cr
& \qquad +{\rm terms\ with\ derivatives} ~.}}
The zero momentum equations of motion of $q$ and $\phi$ are
\eqn\eomphq{\eqalign{&q F_\phi + \phi F_q - \psi_q \psi_\phi=0~, \cr
&2 \phi F_\phi - \psi_\phi^2=0~.}}
(Note that the equation of motion of $\phi$ is basically the same as in \eom.)  The solution of these equations is
\eqn\solcomb{\phi={\psi_\phi^2\over 2F_\phi}~,\qquad q={\psi_q\psi_\phi\over F_\phi}-{\psi_\phi^2\over 2F_\phi^2}F_q~.}
Interestingly, these solutions are independent of high energies data, like $c$ and $\hat c$.

Substituting this solution back into the chiral superfields $\Phi$, $Q$ and denoting the fields in $\Phi$ as in the previous sections, we find
\eqn\newsupi{\Xnl={G^2\over 2F}+\sqrt 2\theta G+\theta^2F~,}
\eqn\newsup{Q_{NL}={\psi_q G \over F}-{G^2\over 2F ^2}F_q+\sqrt 2\theta \psi_q+\theta^2F_q~.}
As before, all the fields in \newsupi\newsup\ are functions of $y^\mu$ \ydefa.  These two superfields satisfy the constraints (which were mentioned in a slightly different context in~\BrignolePE)
\eqn\nilident{\Xnl^2=0~,\qquad Q_{NL}\Xnl=0~.}

By analogy to \Xnlal\ the expression for $Q_{NL}$ can also be written as
\eqn\Qnlal{Q_{NL}= \sqrt 2 \left(\psi_q -{F_q G \over F}\right) \Theta + F_q \Theta^2 ~, \qquad   \Theta =\theta  + {G \over \sqrt 2F}~ .}
This makes it obvious that it satisfies the constraint $X_{NL} Q_{NL} = 0$.  Again, as in~\Xnlal, we find the combination $\Theta=\theta  + {G \over \sqrt 2F}$. Another similarity to \Xnlal\ is that we keep the independent fields $F$ and $F_q$.

Here we have derived the two superfields $\Xnl$ and $Q_{NL}$ from a particular microscopic theory.  To show the uniqueness of the result and the inevitability of the constraint \nilident\ we can follow the procedure outlined around~\XnlSUS, and look for a nonlinear superfield $Q_{NL}$ whose scalar component is expressed in terms of the fields in $\Xnl$ as well as $\psi_q$ (and perhaps $F_q$) such that it transforms properly under supersymmetry.  The expression~\newsup\ is the unique solution of these constraints and \nilident\ follows. Since the constraint is manifestly invariant under supersymmetry we are guaranteed that if it has a nontrivial solution, then it must be consistent with supersymmetry.

The physical interpretation of the solution for $\phi$ was given around \solutionX: $\phi$ does not create a one particle state, but instead, as the superpartner of $\psi_\phi=G$ it creates a two Goldstino state.  Similarly, at low energies $q$ cannot create a one particle squark state.  Instead, as the superpartner of the fermionic operator $\psi_q$ it creates a two particle state -- a quark and a Goldstino.  If $F_q$ is nonzero $q$ can create also a two Goldstino state.

The Lagrangian obtained by substituting the solutions \solcomb\ back into the original Lagrangian is
\eqn\toymatters{\eqalign{
&K=\Xnl \bar\Xnl + Q_{NL} \bar Q_{NL}~, \cr
& W= f\Xnl~.}}
This describes the low energy physics.  As before, we expand in powers of $EM\over F$ and $E\over M$.

It is easy to repeat the analysis around \rocekeq\ and derive additional equations which are related to the equations of motion of \toymatters:
\eqn\additionaleq{\eqalign{
&\Xnl \bar D^2 \bar \Xnl = 4 f \Xnl~,\cr
&\Xnl \bar D^2 \bar Q_{NL} = 0~, \cr
&Q _{NL}\bar D^2 \bar Q_{NL} = 0 ~.}}
These can be used to simplify various higher order corrections to \toymatters.

The leading interaction of the quarks involves two Goldstinos and quarks.  It is easily obtained from substituting the solution for $q$ \solcomb\ into the kinetic term of the squark (equivalently, it follows from the term $Q_{NL}\bar Q_{NL}$ in \toymatters) and we get
\eqn\fourfermions{\CL_{leading~interaction}={1\over f^2}\bar G\ \bar \psi_q\partial^2(G\psi_q)~.}
We see that this term is related to the kinetic term of the quark superfield and therefore it is universal -- it depends only on the SUSY-breaking scale.

As before, we assign $S(\Xnl)=-1$, $S(Q_{NL})=1$.  Then, the terms involving $\Xnl$ in \toymatters\ have $S=0$ and the term with $Q_{NL}$ has $S=4$.  As in section 3, there are positive $S$ corrections which involve only $\Xnl$.  More interesting are other $S=4$ terms which couple $\Xnl$ to $Q_{NL}$ which are similar to \sameScou
\eqn\sameSnl{\int d^4 \theta |\partial X_{NL}|^2 |Q_{NL}|^2 ~.}
It is important that this term is unique.  Other terms with the same fields and two derivatives are equal to it be integration by parts and the constraints.

Clearly, the coefficient of \sameSnl\ is not universal.  It cannot be determined without additional information about the high energy theory.  Conversely, if the high energy theory is known, then the coefficient is calculable.  For example, if the high energy theory is a supersymmetric theory satisfying \MSSMa\ or $m_{soft} \ll \sqrt f$, then the coefficient of \sameSnl\ is calculable; it can be found by integrating out the heavy superpartners and perhaps also the massive gauge bosons.\foot{Actually, in the context of the MSSM as defined in~\lagchi\lagvec, it is easy to prove that~\sameSnl\ is not generated by integrating out the sparticles at tree level. The argument is that the leading order equations of motion of the sparticles amount to replacing all the superfields by their constrained versions (as in the example in subsection 3.2) leading to terms like~\toymatters.  A more careful integration out of the sparticles is carried out by expanding the propagators of the sparticles in momentum, thus adding derivatives while keeping the number of external particles fixed. Therefore, the scaling of the resulting operators is increased. However,~\sameSnl\ has the same scaling as $\int d^4\theta |Q_{NL}|^2$.}

Expanding \sameSnl\ in components we find the coupling
\eqn\fourfermionsi{(\psi_q\sigma_\nu\bar\psi_q)(\partial_\mu G
\sigma^\nu\partial^\mu\bar G)~.}
Unlike the discussion around \sameScou, here this term can lead to quark Goldstino interactions which are of the same scaling as those of \fourfermions.  This follows from the fact that both of them arise from supersymmetric operators with the same scaling $S=4$.  If all the momenta of the scattered Goldstinos and quarks are small and comparable, this term is as important as the contribution from \fourfermions.  However, if the Goldstinos are softer than the quarks (this is impossible in quark-anti-quark scattering), the interaction \fourfermions\ is dominant because in \fourfermionsi\ the derivatives act on the Goldstinos, while \fourfermions\ cannot be brought to this form.

In appendix B we show that even without using the supersymmetry constraints, there are only two operators with this field content and two derivatives.  These are the operators \fourfermions\ and \fourfermionsi.  Now we see that supersymmetry determines the coefficient of \fourfermions\ but the coefficient of \fourfermionsi\ is not determined by low energy considerations.

Note that so far we have only discussed the case of massless quarks. A mass $m$ for the quarks can easily be incorporated by adding $mQ_{NL}^2$ to the superpotential.

Some parts of the discussion of the two Goldstinos and two quarks are very similar to the one in~\refs{\BrignoleFN,\BrignolePE,\BrignoleSK}.  We have presented it here to explain how this easily follows from our formalism and to prepare the ground for subsequent subsections.

\subsec{Complex Scalars}

In the previous subsection we considered the low energy theory of a fermion from a chiral superfield.  Here we will analyze a situation of a complex scalar from a chiral superfield.  Such a situation, where the fermion is much heavier than the scalar, is unnatural and involves fine tuning of the parameters.  Nevertheless, it is worth exploring it for two reasons.  First, we might face a situation where such fine tuning takes place.  Second, this analysis is a good introduction to our discussion in the next subsection which describes real light scalars which can naturally arise if they are Goldstone bosons of some broken global symmetry.

The terminology we will use is that of Higgs physics.  The Higgs chiral superfield $\CH$ includes a complex scalar $H$.  When it is much lighter than its Higgsino superpartner we will derive a nonlinear superfield $\CH_{NL}$.

Based on our experience so far, here and in the later subsections we will use the following procedure. We will postulate a constraint analogous to \nilident, and then solve it.  Then it will be straightforward to check that the solution indeed contains the desired degrees of freedom and hence can be used to write Lagrangians.

We want to eliminate the fermionic component of $\CH$.  This is achieved by imposing
\eqn\strangecons{\Xnl\bar D_{\alphadot}\bar\CH_{NL}=0~,}
or equivalently
\eqn\strangeconsi{\Xnl\bar\CH_{NL}={\rm chiral}~.}
Therefore, $\Xnl$ ``projects'' any function of $\CH_{NL},\bar\CH_{NL}$ on a chiral superfield
\eqn\strangeconsii{\Xnl P(\CH_{NL},\bar\CH_{NL})={\rm chiral}~.}

The constraint \strangeconsi\ is easy to solve in components. The detailed computation is described in appendix C.
The result in components is
\eqn\finalHiggs{\eqalign{
{\cal H}_{NL}= & H+i\sqrt2\theta\sigma^\mu\left(\bar G\over \bar F\right)\partial_\mu H+\theta^2\left(-\partial_\nu\left({\bar G\over \bar F}\right)\bar\sigma^\mu\sigma^{\nu}{\bar G\over \bar F}\partial_\mu H+{1\over 2\bar F^2}\bar G^2\partial^2 H\right) \cr
=& H(\hat y)~,}}
where all the fields in the first line are functions of $y^\mu$ and we have defined
\eqn\newy{\eqalign{
\hat y^\mu =&  y^\mu+i\sqrt 2\theta\sigma^\mu {\bar G(\hat y)  \over \bar F(\hat y)}\cr
= &  y^\mu+i\sqrt 2 \theta\sigma^\mu {\bar G( y) \over \bar F(y)}- 2  \theta \sigma^\mu \partial_\nu \left({\bar G( y) \over \bar F( y)} \right) \theta\sigma^\nu{ \bar G( y)\over \bar F( y)}~.}}
The expression for $\CH_{NL}$ as a function of the independent field $H$ as a function of $\hat y$ is similar to expressions in the literature based on nonlinear realizations~\SamuelUH.  Here we derived it using the constraint \strangeconsi.

As with our previous examples of constrained fields, we see that the superpartner of the scalar $H$ involves an $H$ and a Goldstino.

Using the constraint \strangecons\ and a shift of $X_{NL}$ by a term of the form $\chi X_{NL}$ where $\chi$ is some chiral superfield we find
\eqn\action{\Xnl\bar\Xnl P(\CH_{NL},\bar\CH_{NL})\bigr|_{\theta^4} \sim
\Xnl P(\CH_{NL},\bar\CH_{NL})\bigr|_{\theta^2}+c.c. +{\rm total ~derivatives}~.}
where the last term is supersymmetric because of  \strangeconsii.   These terms lead to an arbitrary scalar potential for $H$.

The superfield \finalHiggs\ does not include the $F_H$-term of the original chiral superfield.  Therefore, terms like $\CH_{NL}^2$ in the superpotential do not give rise to a mass term. Rather, a superpotential for $\CH_{NL}$ leads to terms with Goldstinos.  A general superpotential $W=W(\CH_{NL})$ results in the component Lagrangian
\eqn\supcomii{
W(\CH_{NL})\bigr|_{\theta^2}=  -\pa_\nu\left({\bar G\over\bar F}\right)\bar\sigma^\mu\sigma^\nu{\bar G\over\bar F}\pa_\mu W(H)+{\bar G^2\over 2\bar F^2}\partial^2W(H)
= -{\bar G^2\over 2\bar F^2}\partial^2W(H)+ \dots~,}
where the ellipses represent total derivatives and terms with more Goldstinos.  To show this equality one has to integrate by parts and use the Dirac equation of the Goldstino.   This result will be useful in the next subsection.

\subsec{Real Scalars}

Next we describe a real scalar from a chiral superfield.  We will start with a general solution to this problem and then specialize to a scalar which is a Goldstone boson of a continuous global symmetry.  This symmetry can be an ordinary symmetry or an R-symmetry.

In the previous section we have found a superfield $\CH_{NL}$ describing one complex degrees of freedom and no fermions. Our goal here is to find a consistent way to reduce the number of degrees of freedom to just one real scalar.  We expect to find a superfield of the form $\CH_{NL}$ in \finalHiggs\ but in order to avoid confusion with the Higgs superfield we refer to such a superfield as an axion superfield $\CA_{NL}$.

We want to keep only a single real field $a$ as an independent degree of freedom.  The imaginary part of the bottom component $b$ should be expressed in terms of  $a$ and the components of $\Xnl$.  The constraint we impose is
\eqn\axioncons{\Xnl\left(\CA_{NL}-\bar \CA_{NL}\right)=0~.}
This implies, in particular, the constraint \strangeconsi. Therefore $\CA_{NL}$ should be thought of as a superfield of the form discussed in the previous section with an additional constraint removing one real degree of freedom.  It is easy to solve this equation in components.  The solution is of the form \finalHiggs\ where the bottom component is
\eqn\Haxion{\CA_{NL}\bigr|=a+ib~,}
and $b$ is the following function of the axion $a$ and Goldstinos
\eqn\solutionY{\eqalign{&b={1\over 2} ({G\over F}\sigma^\mu{\bar G\over \bar F})\partial_\mu a+\cdots~,}}
where the ellipses  stand for terms with more fermions and derivatives which are explicitly displayed in appendix D.

Instead of embedding a complex scalar $H$ in a constrained chiral superfield $\CH_{NL}$, we can write $H=a_1 + i a_2$ and embed $a_{1,2}$ in two independent constrained superfields $\CA_{NL1,2}$. So in this sense the superfields $\CH_{NL}$ and $\CA_{NL}$ are on the same footing.
Indeed, the superfield $\CH_{NL}=\CA_{NL1}+i\CA_{NL2}$ satisfies \strangeconsi
\eqn\CHfCA{\Xnl \bar \CH_{NL} = \Xnl(  \bar\CA_{NL1}- i \bar \CA_{NL2})= \Xnl (\CA_{NL1} - i \CA_{NL2}) = {\rm chiral~.}}

The solution \solutionY\ allows us to derive an additional useful identity which follows from $b^3=0$
 \eqn\pureAidentity{ \left(\CA_{NL}-\bar\CA_{NL}\right)^3 =0~.}

When the field ${\cal A}_{NL}$ satisfying \axioncons\ describes a Goldstone boson, the theory is invariant under the spontaneously broken symmetry
\eqn\shiftCA{\CA_{NL} \to \CA_{NL} + \alpha~,}
with real $\alpha$.  Using this symmetry and the identities $\Xnl^2=0$, \axioncons\ and \pureAidentity\ the most general K\"ahler potential without derivatives is
 \eqn\KahlerCA{K= |\Xnl|^2 - {1\over2}f_a^2(\CA_{NL}- \bar\CA_{NL})^2~,}
where the constant $f_a$ is the axion decay constant.

It it is natural to define an exponentiated field with charge one under the broken symmetry
\eqn\eqnexpofield{\CR_{NL}=e^{i{\cal A}_{NL}}~}
and then \axioncons\ implies
\eqn\eqnexpofield{\Xnl\left(|\CR_{NL}|^2-1\right)=0~.}
In terms of $\CR_{NL}$ \KahlerCA\ becomes
 \eqn\KahlerCR{K= |\Xnl|^2 + f_a^2|\CR_{NL}|^2 ~.}
Its simplicity follows from \eqnexpofield\ as well as $\left(|\CR_{NL}|^2-1\right)^3=0$ and $|\CR_{NL}|^2 -{1 \over 4} |\CR_{NL}|^4 = {\rm chiral + antichiral}$.

If $a$ is the Goldstone boson of a broken R-symmetry (which is often being referred to as an R-axion), the transformation law \shiftCA\ should be accompanied by a rotation of $\theta$.  Then $\CR_{NL}$ has R-charge $1$ and we can write the superpotential
\eqn\supaxion{W=f\Xnl+\tilde f\CR_{NL}^2~}
with an arbitrary $\tilde f$.

Using \supcomii\ we see that the term proportional to $\tilde f$ in \supaxion\ leads to the interaction
\eqn\symmbreakingterms{\CL\sim \tilde f \ \bar G^2\partial^2(e^{2ia})\sim \tilde fe^{2ia}\bar G^2(\partial a)^2~.}
If $\tilde f=0$ there is a new ``accidental'' symmetry \shiftCA\ without a $\theta$ rotation. This symmetry acts as $\CR_{NL} \rightarrow e^{i\alpha}\CR_{NL}$ and $\Xnl$ is left invariant. The term $\tilde f$ breaks this symmetry by introducing terms like~\symmbreakingterms. These terms are necessary for the interpretation of $a$ as an R-axion.  Therefore, this term should be added to the action in \ClarkXJ.

In appendix E we present a simple linear theory involving a single chiral superfield which exhibits an R-symmetry and SUSY breaking. After solving the equations of motion of the massive modes we find a single nonlinear chiral superfield which includes both the Goldstino and the axion.  Its action turns out to be identical to the one we find using the two superfields $\Xnl$ and $\CR_{NL}$.

\subsec{Gauge Fields}

In this subsection we discuss the coupling of Goldstinos to low energy gauge fields.  For simplicity, we will restrict our attention to Abelian gauge fields.  The generalization to non-Abelian fields is straightforward.
We will face two distinct issues.  First, we will discuss the gauge invariance of the supersymmetric gauge field which we will need to restrict.  Then we will eliminate the gauginos.

The usual supersymmetric gauge field
\eqn\Vcomp{\eqalign{
V = & \ c+i\theta \chi - i\bar \theta \bar \chi + \theta^2 M + \bar \theta ^2 \bar M -\theta\sigma^m \bar\theta A_m \cr
&+i \theta^2 \bar \theta (\bar \lambda + {i \over 2}\bar \sigma^m \partial_m \chi) - i \bar \theta^2 \theta (\lambda + {i \over 2} \sigma^m \partial_m \bar \chi) + {1 \over 2} \theta^2 \bar\theta^2 (D +{1 \over 2} \partial^2 c) }}
transforms under gauge transformations as
\eqn\gaugetranV{V\to V + \Omega + \bar \Omega}
with chiral $\Omega$.  If our matter system includes charge scalars which are promoted to constrained superfields like $\CH_{NL}$ or $\CR_{NL} = e^{i\CA_{NL}}$, as is the case in possible applications to the Standard Model, we cannot keep the full gauge invariance \gaugetranV. The reason is that the constraints \strangeconsi\ and \axioncons\ are not invariant under \gaugetranV. More concretely, the gauge transformation of $\CA_{NL}$

\eqn\gaugeCA{\CA_{NL} \to \CA_{NL} + i \Omega_{NL}}
is consistent with the constraint \axioncons\ only if $\Omega_{NL} $ is constrained to satisfy
\eqn\gaugeOmegaNL{\Xnl(\Omega_{NL} + \bar \Omega_{NL})=0~.}

This situation is similar to the standard choice of Wess-Zumino gauge.  This gauge choice breaks supersymmetry and the remaining gauge freedom is that of ordinary gauge transformations which are labeled by ${\rm Im}\ \Omega|$.  In our formalism the real scalar degree of freedom in ${\rm Im}\ \Omega|$ is promoted to the nonlinear superfield $\Omega_{NL}$ which satisfies \gaugeOmegaNL.
Therefore, we need to find the analog of Wess-Zumino gauge in our formalism.  It fixes the large group of gauge transformations \gaugetranV\ with arbitrary $\Omega$ and leaves freedom only in $\Omega_{NL}$.  This will have the effect of making the modes $c$, $\chi$ and $M$ of $V$ dependent variables.

The desired gauge choice is
\eqn\constV{\Xnl V_{NL}=0~.}
Its components solution is
\eqn\Vsolc{\eqalign{
c= &{\bar G \bar\sigma^\mu G \over 2|F|^2}A_\mu   + {i\bar G^2 G\lambda\over 2\sqrt2\ \bar F^2 F }  - {i G^2 \bar G \bar\lambda\over  2\sqrt2F^2 \bar F } + {G^2\bar G^2D \over 8|F|^4}  + \dots\cr
\chi=& -i{\bar G\bar \sigma^\mu \over \sqrt2\ \bar F} A_\mu + {\bar G^2\over 2\bar F^2} \lambda + \dots \cr
M=& i{\ \bar G\ \bar\lambda \over \sqrt2 \ \bar F} - {\bar G^2 \over 4\bar F^2} D  + \dots~,}}
where the ellipses represent terms with more Goldstinos and more derivatives.  These terms can be straightforwardly obtained from~\constV; in particular, the series in derivatives and Goldstinos truncates after a finite number of terms.  We will not present them here.

As in \Xnlal\Qnlal\ we can express the gauge field as
\eqn\VNLcomp{\eqalign{
V_{NL}= & -\Theta ^\alpha\bar \Theta^{\dot \alpha} \left[\sigma^m _{\alpha\dot \alpha}  A_m  + { \bar G_{\dot \alpha}G_\alpha D\over |F|^2} + \sqrt 2i \left({\bar G_{\dot \alpha} \lambda_\alpha\over  \bar F} + {G_\alpha \bar \lambda_{\dot \alpha}\over F} \right)\right]\cr
&+i \Theta^2 \bar \Theta \left(\bar \lambda + i  {\bar G D\over \sqrt 2\ \bar F }\right)  - i \bar \Theta^2 \Theta \left(\lambda  - i {G D\over \sqrt 2 F} \right)
 + {1 \over 2} \Theta^2 \bar\Theta^2 D + {\rm terms\ with\ derivatives}~,\cr
\Theta = & \theta +{G \over \sqrt 2 F} ~.}}
Since we neglect terms with derivatives, we can take the arguments of the various fields to be either $x^\mu$ or $y^\mu$ or $\bar y^\mu$.    Using $\Xnl = F\Theta^2$, this expression makes it clear that it satisfies the gauge choice \constV.  It also makes it an obvious analog of the Wess-Zumino gauge; for example, we have $V_{NL}^3=0$.

It is easy to check that the gauge choice \constV\ is invariant only under gauge transformations
\eqn\congt{V_{NL} \to V_{NL} + \Omega_{NL} + \bar \Omega_{NL}~,}
with $\Omega_{NL}$ satisfying \gaugeOmegaNL. Now, with this $V_{NL}$ we can write gauge couplings like $\CR_{NL} e^{V_{NL}} \bar \CR_{NL}$ which are invariant  under \congt.

The second issue we should discuss is the elimination of the gauginos. In many interesting cases, particularly in supersymmetric extensions of the Standard Model, some of the gaugino fields are much heavier than the gauge bosons. It is therefore desirable to have a description without the gauginos to facilitate the study of the low energy effective action.

This is best done in terms of the field strength superfield
\eqn\fs{\eqalign{
W_\alpha=&-i\lambda_\alpha+L_\alpha^\beta\theta_\beta+\sigma^m_{\alpha\alphadot}
\partial_m\bar\lambda^\alphadot \theta^2~,\cr
L_\alpha^\beta =& \delta^\beta_\alpha D-{i\over 2}(\sigma^m\bar\sigma^n)^\beta_\alpha F_{mn}~.}}
A simple constraint which eliminates the gaugino $\lambda$ but leaves the field strength $F_{\mu\nu}$ and the $D$-term is
\eqn\gaugeconstraint{\Xnl W_{\alpha NL}=0~.}

By analogy to \Xnlal\Qnlal\ we can solve \gaugeconstraint\ by writing
\eqn\WNLsola{W_{\alpha NL} = \left(L_\alpha^\beta   -\sqrt 2 \sigma^m _{\alpha \alphadot }\partial_m \bar \lambda^{\alphadot } {G^\beta\over F} \right) \Theta_\beta + \sigma^m _{\alpha  \alphadot}\partial_m  \bar \lambda^{\alphadot}  \Theta^2~,}
which makes it obvious that it satisfies the constraint $X_{NL}W_{\alpha NL}=0$.  Then, in order to match the form \fs\ we need to impose
\eqn\lambdaeqi{-i\lambda_\alpha = L_\alpha^\beta {G_\beta \over \sqrt 2 F}- \sigma^m _{\alpha \alphadot }\partial_m  \bar \lambda^{ \alphadot}  {G^2 \over  2F^2}~.}
It is solved by
\eqn\resultgaugea{\eqalign{
&\lambda=iL\hat G-\hat G^2\partial_m(\bar {\hat G}\bar L)\bar\sigma^m\epsilon -i\hat G^2\sigma^m\bar\sigma^n\partial_m\left(\bar{\hat G}^2\partial_n (L\hat G)\right)-2\hat G^2\bar {\hat G}^2(\partial\hat G)^2\partial_n\bar {\hat G}\bar L\bar\sigma^n\epsilon~, \cr
&\hat G={G\over \sqrt2 F} ~.} }
The combination $L\hat G$ stands for $L^\beta_\alpha \hat G_\beta$ which is the same as $\hat G_\alpha D-{i\over 2}\sigma^m\bar\sigma^n\hat G F_{mn}$ and the combination $\bar {\hat G}\bar L$ is its complex conjugate. As with our other constrained fields, the gaugino is roughly given by the field strength times a Goldstino.

We can use this expression for $\lambda $ in \VNLcomp\ to find
\eqn\VNLcompf{V_{NL}=  -\Theta ^\alpha\bar \Theta^{\dot \alpha} \left(\sigma^m _{\alpha\dot \alpha}  A_m  - { \bar G_{\dot \alpha}G_\alpha D\over |F|^2}\right) + {1 \over 2} \Theta^2 \bar\Theta^2 D + {\rm terms\ with\ derivatives} ~.}
This expression, which we will use below, makes many of the properties of this $V_{NL}$ manifest.

In different physical situations we impose the gauge choice \constV\ and the constraint \gaugeconstraint, or one of them, or neither one.  If we want to eliminate a heavy gaugino we use the constraint \gaugeconstraint.  If we want to restrict the gauge symmetry (we must do it when charged scalars are present) we impose the gauge choice \constV.  For low energy applications in the Standard Model we must impose both \constV\ and \gaugeconstraint.

It is straightforward to write Lagrangians for these superfields. Let us give some examples. Consider a system where the gaugino is heavy. We therefore impose~\gaugeconstraint.
The universal terms in the Lagrangian are obtained, as before, from the kinetic term for the superfield $W_{\alpha NL}$
\eqn\gaugeuni{\int d^2\theta {1\over 4 g^2} W^\alpha_{NL}W_{\alpha NL}~.}
The leading term containing the field strength and Goldstinos is
\eqn\leadingcoupling{{i\over 16 g^2f^2}\left(\bar G_\alphadot F^{\alphadot\dot\beta}\right)\partial_{\beta\dot\beta}\left(F^{\beta\alpha}G_\alpha\right)~.}
Here we used the following notation
$F^{\alpha\beta}=F_{\mu\nu}(\epsilon\sigma^{\mu}\bar\sigma^{\nu})^{\alpha\beta}$ and its complex conjugate.

In section~4 we discussed the operator $\int d^4\theta\ \bigl|X_{NL}\bigr|^2D^\alpha W_{\alpha}$~\lagvec\ and mentioned that it can be removed by a field redefinition~\change. This field redefinition is still possible even if we impose~\constV\ and~\gaugeconstraint.  This easily follows from their invariance under~\change.

Other terms we can consider are coupling to charged fermions and scalars.  The universal terms arise from
\eqn\couplQR{\int d^4 \theta\left( Q_{NL} e^{V_{NL}} \bar  Q_{NL} + \CH_{NL} e^{V_{NL}} \bar  \CH_{NL} + f_a^2 \CR_{NL} e^{V_{NL}} \bar  \CR_{NL}\right)~, }
where the last term represents the Higgsing of the gauge field. Since this theory is everywhere Higgsed, there is no harm introducing a term linear in $V_{NL}$ (see~\KomargodskiPC\ for more details). Therefore, we can also add
\eqn\couplQRi{\int d^4 \theta \xi V_{NL}~.}
This term leads, upon solving the $D$-term equation of motion, to a nonzero $\langle D\rangle$.

We now analyze the coupling of one gauge field to two Goldstinos.  This coupling was first discussed in \LutyNP, and later also, for example, in~\refs{\BrignoleSK,\ClarkAA,\BrignoleCM}. To show an example of how our framework operates, we present here a simple derivation of this coupling and explain its origin and simplicity. We do not need to make any specific assumptions on the detailed way the Standard Model is completed to a supersymmetric theory to derive this coupling.

Focusing on the last term in \couplQR, using the bottom components of $\CR_{NL}$, $\bar\CR_{NL}$ and the vev of $D$ we find
\eqn\vertex{-{f_a^2\langle D \rangle\over 4 f^2}  \bar G\bar\sigma^\mu G A_\mu~.}
Clearly, since it originates from the kinetic term, the coefficient of this operator depends only on universal data: the scale of SUSY breaking $f$, the scale of gauge symmetry breaking $f_a$ and the value of the $D$-term which is described by a term in the low energy effective Lagrangian. As in subsection 5.1, terms that have the same scaling as the universal one $|\CR_{NL}|^2 e^{V_{NL}}$, for example $|\partial X_{NL}|^2 |\CR_{NL}|^2 e^{V_{NL}}$, arise at subleading orders and are negligible.\foot{Another analogy with our discussion in subsection~5.1 is the fact that $|\partial X_{NL}|^2 |\CR_{NL}|^2 e^{V_{NL}}$ does not arise by integrating out the sparticles at tree level. This is due to the same reason as in subsection 5.1.}

We would like to make a few comments about \vertex\ which are important in comparisons with the literature:
\item{1.} Using integration by parts and the equations of motion of $G$ and $A_\mu$ the coupling~\vertex\ can also be written as
    \eqn\GGFC{\bar G \bar \sigma^\mu G A_\mu \qquad \sim \qquad \bar G\bar\sigma^\nu G \partial^\mu F_{\mu\nu}+ \dots\qquad \sim  \qquad \partial^\mu\bar G\bar\sigma^\nu(\partial_\mu G) A_\nu + \dots~.}
\item{2.} Given that the gauge field is massive with $m_A \sim g f_a$, it can be integrated out and lead to additional couplings of Goldstinos to quarks \eqn\qqGGLP{\sim {\langle D\rangle\over f^2 m^2_A}\partial^\mu\bar G \bar\sigma^\nu(\partial_\mu G) \bar \psi_q\bar\sigma_\nu \psi_q ~,}
    where $\psi_q$ is some matter fermion.  This is of the form of the non-universal four fermion term~\fourfermionsi\ which corresponds to the superspace expression~\sameSnl.  As a check, note that the universal term \toymatters\fourfermions\ is indeed not corrected.  This contribution to the non-universal four fermion term is comparable to that of~\fourfermions\ and its magnitude is predicted by the MSSM. Our discussion here is in agreement with~\BrignoleCM.
\item{3.} In appendix F we calculate this term in the context of the MSSM and we reproduce the answer~\vertex\ with $f_a ^2 = v_u^2 + v_d^2$, $\langle D\rangle = -{g^2\over 2}\left(v_u^2-v_d^2\right)$. This computation demonstrates the simplicity of our method compared to explicit microscopic calculations. The latter also obscure the universality of the answer. (For example, see how the gaugino mass dependence drops out in appendix F.)
\medskip
The conclusion of this subsection is that we derived~\vertex\ from a universal low energy term in the Lagrangian~\couplQR. This explains its simplicity and generality. Without limiting ourselves to any specific framework we see that~\vertex\ (and~\qqGGLP) arises in a universal form whenever there is a Higgsed gauge symmetry.

\bigskip
\centerline{\bf Acknowledgements}
We would like to thank N.~Arkani-Hamed, J.~Bagger, M.~Dine, J.~Distler, J.~Maldacena, P.~Meade, M.~Papucci, M.~Peskin, M.~Rocek, A.~Schwimmer, D.~Shih, Y.~Tachikawa, S.~Thomas and S.~Weinberg for useful discussions. The work of ZK was supported in part by NSF grant PHY-0503584 and that of NS was supported in part by DOE grant DE-FG02-90ER40542. Any opinions, findings, and conclusions or recommendations expressed in this material are those of the author(s) and do not necessarily reflect the views of the funding agencies.

\appendix{A} {Uniqueness of Leading Scaling Action}

In this appendix we prove that the most general Lagrangian with $S=0$ can always be written after suitable changes of variables as
\eqn\AVactionapp{{\cal L}=\int d^4\theta X_{NL}\bar X_{NL} +\int d^2\theta fX_{NL}+\int d^2\bar \theta   f  \bar X_{NL}~.}

Using $X_{NL}^2=0$, the list of superfields with negative $S$ is
 \eqn\negsca{\Xnl~, \bar\Xnl~, D_\alpha\Xnl~,\bar D_{\alphadot}\Xnl~.}
In addition, there are some superfields with $S=0$
\eqn\zerosca{D^2\Xnl~, \bar D^2\bar\Xnl~, \partial_\mu\Xnl~, \partial_\mu\bar\Xnl~.}
Lastly, there are many operators with positive $S$. We will not need to specify these operators here.

First, we observe that there are no possible superfields with $S<-2$. This follows because all the operators with negative $S$ have fermions in their bottom components and therefore we cannot multiply too many of them.
This means that the most general way to generate $S=-2$ operators (which become $S=0$ operators after a $\int d^4\theta$ integral) is to combine operators from \negsca\ to have $S=-2$ and multiply by some function of the $S=0$ operators in \zerosca. Note however that $\partial \Xnl, \partial\bar\Xnl$ have at least one fermion in their bottom component and therefore they cannot be used. Therefore, we end up with three different possible terms in the K\"ahler potential
\eqn\threepos{\eqalign{& \Xnl\bar \Xnl P_1(D^2\Xnl,\bar D^2\bar\Xnl)~,\cr
& \Xnl \left(\bar D\ \bar\Xnl\right)^2 P_2(D^2\Xnl,\bar D^2\bar\Xnl)+c.c~,\cr&
\left( D \Xnl\right)^2\left(\bar D\ \bar\Xnl\right)^2 P_3(D^2\Xnl,\bar D^2\bar\Xnl)~.}}
The first term in \threepos\ can always be reduced to $|\Xnl|^2$ by successive uses of $X\bar D^2\bar\Xnl=4f \Xnl$. For the second term we apply
$\bar\Xnl\bar D_{\alphadot} \bar\Xnl=0$ (which follows from $\bar\Xnl^2=0$) to find $\left(\bar D\ \bar\Xnl\right)^2\sim\bar\Xnl\bar D^2\bar\Xnl$. Hence, it reduces to a term of the first type. Similarly, the last term in \threepos\ can be reduced to a term of the first type, which can be transformed to $|\Xnl|^2$. Finally, similar reasoning applies to the superpotential and hence~\AVactionapp\ is the most general $S=0$ Lagrangian.

\appendix{B}{Terms with Four Fermions and Two Derivatives}

Our purpose in this appendix is to understand the general possible couplings of two massless fermions and two Goldstinos.  We will impose Lorentz invariance and the Dirac equation.  We will not use any constraints from supersymmetry.

A priori, two fermions $\chi $ and $\psi$ and two derivatives can be coupled as (all the operators with two derivatives acting on the same fermion can be removed by integration by parts)
\eqn\listofop{\eqalign{&A_1=(\partial\psi)\partial(\bar\psi)\chi\bar\chi~,\qquad A_2=\psi\bar\psi\partial(\chi)\partial(\bar\chi)~, \qquad A_3=\partial(\psi)\bar\psi\partial(\chi)\bar\chi~,\qquad A_4=\partial(\psi)\bar\psi\chi\partial(\bar\chi)~, \cr & A_5=\psi\partial(\bar\psi)\partial(\chi)\bar\chi~,\qquad A_6=\psi\partial(\bar\psi)\chi\partial(\bar\chi)~.}}
Note that $A_3^* = A_6$.  Using the fact that the Dirac equation projects out the spin $1 \over 2$ component of $\partial \psi $ and $\partial \chi$, there are two independent Lorentz contractions in $A_3$ and $A_6$ and a single contraction in the other operators:
\eqn\listofopii{\eqalign{&A_1=(\partial_\mu\psi)\partial^\mu(\bar\psi)\chi\bar\chi~,\qquad A_2=\psi\bar\psi\partial_\mu(\chi)\partial^\mu(\bar\chi)~,\qquad A_4=\partial_\mu(\psi)\bar\psi\chi\partial^\mu(\bar\chi)~,\cr&  A_5=\psi\partial_\mu(\bar\psi)\partial^\mu(\chi)\bar\chi~.}}

Using integration by parts and $\partial^2 \psi = \partial^2 \chi =0 $ we can remove $A_2$, $A_4$ and $A_5$ from our list of independent operators. So, we remain with
\eqn\listofopiii{\eqalign{&A_1=\partial_\mu(\psi)\partial^\mu(\bar\psi)\chi\bar\chi~, \qquad A_3=\partial(\psi)\bar\psi\partial(\chi)\bar\chi~,\qquad A_6=\psi\partial(\bar\psi)\chi\partial(\bar\chi)~.}}

This basis can be further restricted using
 \eqn\lemmaa{\partial_\mu\chi^\alpha\sigma_{\alpha\dot\alpha}^\nu\sigma^\mu_{\beta\dot\beta}
\partial_\nu\bar\chi^{\dot\beta}=2\partial^\mu\chi_\beta\partial_\mu
\bar\chi_{\dot\alpha}~, }
which can be proven as follows
\eqn\proofa{\eqalign{\partial_\mu\chi^\alpha\sigma_{\alpha\dot\alpha}^\nu
\sigma^\mu_{\beta\dot\beta}\partial_\nu\bar\chi^{\dot\beta} & =
\partial_\mu\chi^\alpha\partial_\nu\bar\chi^{\dot\beta}
\left(\sigma_{\alpha\dot\alpha}^\nu\sigma^\mu_{\beta\dot
\beta}+\sigma_{\alpha\dot\alpha}^\mu\sigma^\nu_{\beta\dot\beta}\right)\cr&=
\partial_\mu\chi^\alpha\partial_\nu\bar\chi^{\dot\beta}\left(-\eta^{\mu\nu}
\epsilon_{\alpha\beta}\epsilon_{\dot\alpha\dot\beta}+4(\sigma^{\rho\mu}
\epsilon)_{\alpha\beta}
(\epsilon\bar\sigma^{\rho\nu})_{\dot\alpha\dot\beta}\right)\cr&=
\partial^\mu\chi_\beta\partial_\mu\bar\chi_{\dot\alpha}+
4\partial_\mu\chi^\alpha(\sigma^{\rho\mu}\epsilon)_{\alpha\beta}
\partial_\nu\bar\chi^{\dot\beta}
(\epsilon\bar\sigma^{\rho\nu})_{\dot\alpha\dot\beta}\cr
&=\partial^\mu\chi_\beta\partial_\mu\bar\chi_{\dot\alpha} -{1\over 4}\partial_\mu\chi^\alpha(\sigma^{\rho}\bar\sigma^{\mu}\epsilon)_{\alpha\beta}
\partial_\nu\bar\chi^{\dot\beta}
(\epsilon\bar\sigma^{\nu}\sigma^{\rho})_{\dot\alpha\dot\beta} \cr
&=\partial^\mu\chi_\beta\partial_\mu\bar\chi_{\dot\alpha} +{1\over 2}\partial_\mu\chi^\alpha\sigma_{\alpha\dot\alpha}^\nu\sigma^\mu_{\beta\dot\beta}
\partial_\nu\bar\chi^{\dot\beta }.}}

Using \lemmaa\ and the Dirac equation and integration by parts $A_1$ is of the form $A_3$.  Since $A_3^*=A_6$, we end up with the two dimensional space of operators of the form $A_3$. Writing these two operators explicitly one finds that they are both manifestly real (after integration by parts).
In the text we preferred to use a different basis \fourfermions \fourfermionsi\ for the same two dimensional space of operators.

\appendix{C}{The Nonlinear Higgs Superfield}

Here we solve the (complex conjugate of the) constraint \strangecons. In the $y$ coordinate we can write the Higgs superfield as
\eqn\anchHiggs{\CH_{NL}= H(y)+\sqrt2\theta\psi_H(y)+\theta^2 F_H(y)~.}
Let us now write the same superfield in the coordinate $\bar y$. This is easily done remembering that $y^m=\bar y^{ m}+2i\theta\sigma^m\bar \theta$. We therefore expand and get
\eqn\anchHiggs{\eqalign{&\CH_{NL}= H(\bar y^{ m}+2i\theta\sigma^m\bar \theta)+\sqrt2\theta\psi_H(\bar y^{ m}+2i\theta\sigma^m\bar \theta)+\theta^2 F_H(\bar y^{ m}+2i\theta\sigma^m\bar \theta)\cr
&=H+\sqrt2\theta\psi_H+\theta^2 F_H+2i\theta\sigma^m\bar\theta\pa_m H+\theta^2\bar\theta^2\partial^2 H-i\sqrt2\theta^2\pa_m\psi_H\sigma^m\bar\theta~,}}
where all the fields in the second line are functions of $\bar y^m$. Multiplying it with
\eqn\antiGoldstino{\bar\Xnl={\bar G^2\over 2\bar F}+\sqrt2\bar\theta  \bar G +\bar\theta^2\bar F~,}
where again all the fields are functions of $\bar y^m$, we should get an antichiral superfield. This is easily imposed by requiring that all the components but $1,\bar\theta,\bar\theta^2$ cancel.  They lead after some algebra to:
\eqn\compii{\eqalign{
\psi_H= & i\sigma^m{\bar G\over \bar F}\pa_m H~, \cr
F_H= &-\pa_m\left({\bar G\over\bar F}\right)\bar\sigma^n\sigma^m{\bar G\over\bar F}\pa_nH+{\bar G^2\over 2\bar F^2}\partial^2H~.}}

\appendix{D}{The Axion Superfield}

We present here the complete solution to the constraint
\eqn\axionconsi{\Xnl\left(\CA_{NL}-\bar\CA_{NL}\right)=0~.}
From the bottom component we get
\eqn\axioncomp{\eqalign{ bG^2=0~.}}
This is solved by assuming that $b$ begins with two fermions.
The equation of the $\theta^2$ component is
\eqn\axioncompii{2b-({G\over F}\sigma^\mu{\bar G\over \bar F})\partial_\mu H+i{G^2\over 2F^2}\left(\partial_\nu\left({\bar G\over \bar F}\right)\bar\sigma^\mu\sigma^{\nu}{\bar G\over \bar F}\partial_\mu H-{1\over 2\bar F^2}\bar G^2\partial^2 H
\right)=0~.}

Since we know from~\axioncomp\ that $b$ includes fermions, we can solve the equations iteratively
\eqn\solutionY{\eqalign{&b={1\over 2} ({ G\over F}\sigma^\mu{\bar G\over \bar F})\partial_\mu a-\left({i\over 8}{ G^2\over F^2}\partial_\nu\left({\bar G\over \bar F}\right)\bar\sigma^\mu\sigma^\nu{\bar G\over\bar F}\partial_\mu a+c.c.\right)\cr&-
{ G^2\bar G^2\over 32F^2\bar F^2}\partial_\mu\left(\bar G\over \bar F\right)\left(\bar\sigma^\rho\sigma^\mu\bar\sigma^\nu+\bar\sigma^\mu\sigma^\nu
\bar\sigma^\rho\right)\partial_\nu\left( G\over F\right)\partial_\rho a~.}}
The first term in this expansion was quoted in subsection 5.4.

\appendix{E}{A Microscopic Theory for SUSY and R-Symmetry Breaking}

Consider an R-invariant theory based on a single chiral superfield field with a general K\"ahler potential and a linear superpotential
\eqn\Rbreakingmodel{\eqalign{&K( \Phi\overline{ \Phi})=
 \Phi\overline{ \Phi}+{c_1\over M^2}\left( \Phi\overline{ \Phi}\right)^2+{c_2\over M^4}\left( \Phi\overline{ \Phi}\right)^3+\cdots~,\cr&
W=f  \Phi~.}}
Here the ellipses represent higher order terms. The R-charge of $\Phi$ is fixed to be~$2$. For certain choices of parameters the theory has a classical vacuum at nonzero $\Phi$, thereby spontaneously breaking both supersymmetry and the R-symmetry. In this case we expect to find at low energies a Goldstino and an R-axion.

Here the Goldstino and the R-axion originate from the same superfield, which becomes a nonlinear superfield, $\Phi_{NL}$.  Denoting the R-axion by $a$
we expect to find $e^{2ia}$ in the bottom component of $\Phi_{NL}$ and the Goldstino in its $\theta$ component. This motives us to try to express $\Phi_{NL}$ as a linear combination of our constrained superfields $X_{NL}$ and $\CR_{NL}$.

Normalizing $ \Phi$ conveniently, we can write
\eqn\linearcom{ \Phi_{NL}=\CR_{NL}^2+\alpha\Xnl~,}
where $\alpha$ is some complex number (with mass dimension $-1$). Using the relations in subsection 5.4 it is easy to see that
\eqn\cbic{\left(| \Phi_{NL}|^2-1\right)^3=0~.}
In addition, one can check that up to total derivatives $| \Phi_{NL}|^4 \sim | \Phi_{NL}|^2$. Hence, the most general theory without covariant derivatives is
\eqn\sinfieldaxion{{\cal L}=f_a^2 \Phi_{NL}\overline{ \Phi}_{NL}
\bigr|_{\theta^4}+\left(\hat f \Phi_{NL}|_{\theta^2}+c.c\right)~.}
Of course, here $\hat f$ has dimension 3.
The physical decay constant of the axion is $2f_a$ and the SUSY breaking scale is $f={\hat f\over f_a}$.

Rewriting this theory in terms of $X_{NL},\CR_{NL}$ we find a Lagrangian of the form
\eqn\diction{{\cal L}=\left(4f_a^2|\CR_{NL}|^2+f_a^2\alpha^2|\Xnl|^2\right)\bigr|_{\theta^4}+\biggl[\left( \hat f\alpha\Xnl+\hat f\CR_{NL}\right)\bigr|_{\theta^2}+c.c\biggr]~.}
We can conveniently choose $\alpha=f_a^{-1}$ and obtain
\eqn\diction{{\cal L}=\left(4f_a^2|\CR_{NL}|^2+|\Xnl|^2\right)\bigr|_{\theta^4}+\biggl[\left( {\hat f\over f_a}\Xnl+\hat f\CR_{NL}\right)\bigr|_{\theta^2}+c.c\biggr]~.}
We explicitly see that there are only two adjustable parameters, while the most general case in subsection 5.4 has an additional parameter. The remaining parameter of the theory of a Goldstino and an R-axion can be recovered by adding a higher derivative term such as $\int d^4\theta \bar\Phi_{NL}(D\Phi_{NL})^2+c.c.$.

\appendix{F}{Massive Gauge Boson Decay}

Here we consider a microscopic MSSM-like theory which leads to the coupling $G\bar\sigma\bar GA_\mu$~\vertex.  We start with a supersymmetric $U(1)$ gauge theory with two charged superfields $H_u$, $H_d$ with charges $1$, $-1$ respectively and add to it soft terms at the scale $m_{soft}$.  As in section 4 we take $m_{soft} \ll \sqrt f$.

We start with the Lagrangian
\eqn\deflag{\eqalign{&\int d^2\theta \left(\left({1\over 4g^2}+{m_{\tilde g}\over 2f}X_{NL}\right)W_\alpha^2+\mu H_uH_d \right)+c.c.\cr&+
\int d^4\theta\left(\bar H_ue^VH_u-{m_u^2\over f^2}|X_{NL}H_u|^2e^V+\bar H_de^{-V}H_d-{m_d^2\over f^2}|X_{NL}H_d|^2e^{-V}\right)~. }}
As in section 4, a microscopic $D$-term can be shifted away.

Clearly, classical integration out of massive particles cannot lead to the three particle coupling $G\bar\sigma\bar GA_\mu$.  Instead, it can be generated from off-diagonal terms in the fermions mass matrix.  For that, we focus on the zero-momentum terms which are at most quadratic in fermions and we replace the Higgs fields by their VEVs ($v_{u,d}$) everywhere. The contributions from the usual supersymmetric terms are
\eqn\deflagi{\eqalign{\CL_{SUSY}= & {1\over 2g^2} D^2+|F_u|^2+|F_d|^2+\left(\mu F_u v_d+\mu F_d v_u-\mu\psi_u\psi_d+c.c.\right)+{D\over 2}\left(v_u^2-v_d^2\right)\cr&
+{i\over \sqrt2}\left(v_u\psi_u\lambda-v_d\psi_d\lambda-c.c.\right)+\left({1\over2}\bar \psi_u\bar\sigma^\mu\psi_u-{1\over 2}\bar\psi_d\bar\sigma^\mu\psi_d\right)A_\mu~. }}
The leading order couplings to $\Xnl$ lead to
\eqn\deflagii{\eqalign{\CL_{G}= & {v_um_u^2\over f}(\psi_uG+c.c.)+{v_dm_d^2\over f}(\psi_dG+c.c.)-\left({m_u^2v_u^2\over 2f^2}\bar G\bar\sigma^\mu G-{m_d^2v_d^2\over 2f^2}\bar G\bar\sigma^\mu G\right)A_\mu\cr &-{1\over2}m_{\tilde g}\lambda^2+{im_{\tilde g}\over \sqrt2 f}D\lambda G~.
}}

The equations of motion of the Higgs fields lead to
\eqn\eomi{\eqalign{&\mu^2 +m_u^2-{D\over 2}=0~, \cr
&\mu^2 +m_d^2+{D\over 2}=0~,}}
and those of the the massive fermions $\psi_{u,d},\lambda$ are
\eqn\eomfi{\eqalign{
&\mu\psi_u+{i\over \sqrt2}v_d\lambda-{v_dm_d^2\over f}G=0~,\cr
&\mu\psi_d-{i\over \sqrt2}v_u\lambda-{v_um_u^2\over f}G=0~, \cr
&m_{\tilde g}\lambda-{im_{\tilde g}\over \sqrt2 f}D G-{i\over \sqrt2}v_u\psi_u+{i\over \sqrt2}v_d\psi_d=0~.}}
The solution of these equations is
\eqn\eomfiii{\lambda={iD G\over \sqrt2 f}~,\qquad \psi_u=-{\mu v_d\over f}G~,\qquad \psi_d=-{\mu v_u\over f}G~. }
Note that the dependence on $m_{\tilde g}$ has completely disappeared.

Finally, we substitute this solution back into the last term in the Lagrangian~\deflagi\ and combine it with the term coupling the Goldstino to the vector field in \deflagii\ to find
\eqn\op{-{\langle D \rangle\over 4f^2}\left(v_d^2+ v_u^2\right)\bar G\bar\sigma^\mu GA_\mu~.}
Here $\langle D\rangle = -{g^2\over 2}\left(v_u^2-v_d^2\right)$.
The result of this simple calculation is in agreement with~\vertex\ and~\LutyNP.

\listrefs
 \end